\documentclass[12pt]{article}

\usepackage{amsmath}
\usepackage{amsfonts}
\usepackage{amssymb}
\usepackage{array}
\usepackage{upgreek}
\usepackage{longtable}
\usepackage{booktabs}
\usepackage[toc,page]{appendix}
\usepackage{caption}
\usepackage{subcaption}
\usepackage{caption}
\usepackage{graphicx}
\usepackage{tikz}
\usepackage{multicol}
\usepackage{chngpage}

\usepackage[backend=biber, natbib=true,citestyle= authoryear-comp,sorting=ynt,maxcitenames=4]{biblatex}
\usepackage[colorlinks=true,linkcolor=blue,citecolor=blue]{hyperref}
\DeclareFieldFormat{citehyperref}{%
  \DeclareFieldAlias{bibhyperref}{noformat}% Avoid nested links
  \bibhyperref{#1}}

\savebibmacro{cite}
\savebibmacro{parencite}

\renewbibmacro*{cite}{%
  \printtext[citehyperref]{%
    \restorebibmacro{cite}%
    \usebibmacro{cite}}}

\renewbibmacro*{parencite}{%
  \printtext[citehyperref]{%
    \restorebibmacro{parencite}%
    \usebibmacro{parencite}}}

\begin{document}

\begin{center}
{\Large
	{\sc  Clustering Longitudinal Ordinal Data via Finite Mixture of Matrix-Variate Distributions}
}
\bigskip

Francesco Amato $^{1}$, Julien Jacques $^{1}$, Isabelle Prim-Allaz $^{2}$
\bigskip

{\it
$^{1}$ Univ Lyon, Univ Lyon 2, ERIC, Lyon.\\
$^{2}$ Univ Lyon, Univ Lyon 2, COACTIS, Lyon.
\{\text{francesco.amato, julien.jacques, isabelle.prim-allaz}\}@univ-lyon2.fr\\ 
}
\end{center}
\bigskip

%--------------------------------------------------------------------------

\medskip

{\bf Abstract.}In social sciences, studies are often based on questionnaires asking participants to express ordered responses several times over a study period. We present a model-based clustering algorithm for such longitudinal ordinal data. Assuming that an ordinal variable is the discretization of an underlying latent continuous variable, the model relies on a mixture of matrix-variate normal distributions, accounting simultaneously for within- and between-time dependence structures. The model is thus able to concurrently model the heterogeneity, the association among the responses and the temporal dependence structure. An EM algorithm is developed and presented for parameters estimation, and approaches to deal with some arising computational challenges are outlined. An evaluation of the model through synthetic data shows its estimation abilities and its advantages when compared to competitors. A real-world application concerning changes in eating behaviors during the Covid-19 pandemic period in France will be presented.\\


\bigskip\bigskip

%--------------------------------------------------------------------------

\section{Context}
In many areas of humanities and social sciences, the studies are based on questionnaires. The most common kind of questions, and therefore collected data, are ordinal, as for instance in marketing studies where people are asked to evaluate some products or services on an ordinal scale \citep{dillon1994marketing}. Ordinal data occur when the categories are ordered \citep{Agresti2010Apr}. Ordinality is a characteristic of the meaning of measurements \citep{Stevens1946Jun}, and distinct levels of an ordinal variable differ in degree of dissimilarity more than in quantity \citep{Agresti2010Apr}.

Often, these questionnaires are completed by participants several times over the study period. The researchers then analyse these questionnaires to determine typical behaviours within the studied population, being especially interested in their time evolution. Nonetheless, modelling temporal evolution is far from trivial. The most basic approach consists in performing analyses independently at each temporal phase, and then trying \textit{a posteriori} to find links between these different analyses, by seeking from one phase to the other to find similar or different typical behaviours. An example is \cite{quality}, clustering of ordinal data for an application in psychology.  The ideal way to cluster temporal data would be to account for the temporal evolution, modelling all the responses to the questionnaires at the same time. We propose a model-based clustering technique aiming at facilitate such temporal analysis, by grouping together the units behaving similarly in time.

%\subsection{Model-based clustering} 
Over the decades, research has produced a vast number of different approaches to clustering. From our prospective, probabilistic (or model-based) clustering offers the advantage of clearly stating the assumptions behind the clustering algorithm, and allows cluster analysis to benefit from the inferential framework of statistics to address some of the practical questions arising when performing clustering: determine the number of clusters, detecting and treating outliers, assessing uncertainty in the clustering \citep{bouveyron2019model}.

%\subsection{Our idea} 
Our model proposes to cluster all the ordinal responses at the same time, grouping together the units behaving similarly in time. Moreover, it also aims at being easily understandable and interpretable by practitioners with non-statistical background.

\subsection{Related works}
\label{sec:relworks}

% ORDINAL/MIXED DATA
%A classic strategy currently used for the analysis of questionnaires consists of independently analysing the questionnaires of each time phase, and this by standardizing the answers to the questions so that they are all of the same type: either all categorical nominal or all quantitative. 
Although ordinal data are certainly the type most encountered in questionnaires, they are either transformed according to a Likert scale \citep{likert} into quantitative data \citep{Lewis2005Mar}, or transformed into nominal data by ignoring the order \citep{latentgold}.
In the first case, even if there is a whole literature on the construction of Likert scales, the introduction of a notion of distance between categories necessarily brings a bias in the analysis \citep{Liddell2018Nov}. In the second case, less often used nevertheless, one loses essential information by not taking into account the notion of order within the categories.\\
Ordinal data do not have metric information. One classical model to treat ordinal data as in a ordinal-scale model are the traditional \textit{ordered-probit} models \citep{McKelvey1975Jan,winship1984regression,becker1992probit}. This model describes the probability of a ordinal response as the cumulative normal probability between two thresholds on an underlying latent continuous distribution, generally chosen to be Gaussian. This model is generally regarded as one of the standards in both frequentist and Bayesian frameworks \citep{lynch2007bayes,Kruschke2015}.\\
More recently, other approaches to deal with such kind of data has been developed. In the clustering context we are interested in, the examples spans from \cite{Elia2005Jun}, that introduces the CUB model, later developed through the R package \texttt{CUB} \citep{Iannario2016Jun},
 to \cite{Giordan2011Mar} and more recently \cite{Ranalli2016Jan}; \cite{Fernandez2016Jan}. In a co-clustering context, the R package \texttt{ordinalClust} \citep{ordinalclust} makes use of the BOS (Binary Ordinal Search) distribution introduced by \cite{Biernacki2016Sep} and extended for co-clustering by \cite{jacques2018model}. A
mixture of item response models was developed to
for ordinal response data in the Bayesian framework by \cite{mcparland2013clustering}, to be later expanded in the frequentist paradigm and to handle mixed data in \cite{clustmd}. More recently, \cite{bouveyron2020ordinal} proposed a new model that relies on latent continuous random variables to perform co-clustering.

% LONGITUDINAL DATA

% De la Cruz-Mes´ıa, Quintana & Marshall (2008) use a mixture of nonlinear hierarchical models. The modelling paradigm that they propose, which is essentially an extension of Pauler & Laird (2000), makes each component density subject-specific and the only modelling of the component covariance matrix that they engage in is the imposition of the isotropic constraint.

% The most common formulation for modeling the temporal correlation in longitudinal data consists of introducing continuous time-varying individual random effects that follow an autoregressive latent model of order 1,

Similarly, several approaches to clustering longitudinal data were developed. In \cite{Brendan2010} the authors developed a model-based clustering framework for longitudinal continuous data by using Gaussian mixture models and applying the modified Cholesky decomposition to the group covariance matrices. Doing this, the new derived elements can be interpreted as generalized auto-regressive parameters and innovation variances. Moreover, a series of possible constraints are presented in order to give rise to more parsimonious models. In the context of  generalized linear latent variable models (GLLVMs), \cite{cagnone2018multivariate} introduced a methodological framework that includes two levels of latent variables: one continuous hidden variable for dimension reduction and clustering and a discrete random variable accounting for
the dynamics modelled through a latent Markov model. In the R package \texttt{mixAK} \citep{mixAK} the basis for clustering is a mixture of multivariate generalized linear mixed models. In \cite{Vavra2023Jun} a mixture distribution is additionally assumed for random effects.

An other approach to clustering longitudinal data consists in arranging the data in a three-way format and modelling them through a matrix-variate mixture model. This approach offers the advantage of accounting for the overall time-behavior, grouping together the units that have a similar pattern across and within time. While not being new \citep{basford1985mixture}, matrix-variate distributions have recently gained attention, and mixtures of matrix-normals (MMN) have been developed and applied both in a frequentist framework in \cite{Viroli2011Oct} and within a Bayesian one by \cite{Viroli2011Dec}, where it was used to cluster Italian provinces based on a longitudinal crime-related score. From a frequentist point of view, these models represent a natural extension of the multivariate normal mixtures to account for temporal (or even spatial) dependencies, and have the advantage of being also relatively easy to estimate by means of EM algorithm (a nice short description of the EM application to MNN is provided in §2.1 of \cite{wang2020variable}). 
\cite{Anderlucci2015Jun} extends on the work of \cite{Brendan2010} and incorporates the idea of the modified Cholesky decomposition in the matrix-variate regression model developed by \cite{matrixregression}, elaborating a family of more parsimonious models. 
More recently, in \cite{dougru2016finite,gallaugher2018finite} and \cite{Melnykov2018Sep,melnykov2019studying} extensions for non-normal skewed matrix-variate mixture model have been proposed and applied. %However, matrix-variate models suffer from over-parametrization that leads to estimation issues. 
An attempt to generalize the class of parsimonious models derived by the decomposition of the covariance matrices in a mixture of matrix-normal model has been carried out \citep{Sarkar2020Feb}. A new comprehensive R package to apply this family to clustering continuous three-way data \citep{Zhu2022Mar} has been proposed, endeavoring the creation of a \texttt{mclust} \citep{mclust} for three-way continuous data.

\subsection{The idea}
\label{sec:idea}
As we aims at develop a model easily understandable and interpretable by practitioners with non-statistical background, we found matrix-variate distributions particularly fit, as shown in \cite{Alaimo2023Jan}. Moreover, as noticed in \cite{Anderlucci2015Jun}, the use of matrix-variate distributions allow to drop the conditional independence assumption, frequently implied in longitudinal latent variable models. 

Despite the efficacy of matrix-variate distributions, up to now these methods have only been applied to continuous data. We introduce a Mixture for Ordinal Matrices (MOM) model, aiming at expanding  the use to matrix-variate mixtures to ordinal data in an unsupervised learning context.

In the following Sections \ref{sec:model} and \ref{sec:inference} we will detail our model and the EM algorithm to perform inference. In Section \ref{sec:evaluation} the results on synthetic data are presented to assess the performance of the model. Finally, in Section \ref{sec:real} an application on real data concerning grocery shopping preferences by a French sample during the Covid-19 pandemic period is outlined.

\section{Model}
\label{sec:model}
\subsection{Preliminaries}
Let $Z \sim \mathcal{MN}_{(J\times T)}(M,\Phi,\Sigma)$, that is a matrix-variate normal distribution where $M \in \mathbb{R}^{J\times T}$ is the matrix of means, $\Phi \in \mathbb{R}^{T \times T}$ is a covariance matrix containing the variances and covariances between the $T$ occasions or times and $\Sigma \in \mathbb{R}^{J \times J}$ is the covariance matrix containing the variance and covariances of the $J$ variables. The matrix-normal probability density function (pdf) is given by
\begin{equation}
f(Z|M,\Phi,\Sigma) = (2\pi)^{-\frac{TJ}{2}}|\Phi|^{-\frac{J}{2}}|\Sigma|^{-\frac{T}{2}}
\exp\left\{-\frac{1}{2} \text{tr}[ \Sigma^{-1}(Z-M)\Phi^{-1}(Z-M)^{\intercal}] \right\}.
\end{equation}
The matrix-normal distribution represents a natural extension of the multivariate normal distribution, since if $Z \sim \mathcal{MN}_{(J\times T)}(M,\Phi,\Sigma)$, then $\text{vec}(Z) \sim \mathcal{MVN}_{JT}(\text{vec}(M),\Phi \otimes \Sigma)$, where $\text{vec}(.)$ is the vectorization operator and $\otimes$ denotes the Kronecker product. The property of rewriting the general covariance  matrix $\Psi \in \mathbb{R}^{JT \times TJ}$ as $\Psi=\Phi \otimes \Sigma$ is called separability condition. Then, the  mean and the variance of the matrix-normal distribution
are:
\begin{align}
\label{props}
    \mathbb{E}(\text{vec}(Z)|M,\Phi,\Sigma)=\text{vec}(M)\quad \text{ and } \quad
    \mathbb{V}(\text{vec}(Z)|M,\Phi,\Sigma)= \Sigma \otimes \Phi.
\end{align}
Being a special case of the multivariate normal distribution, the matrix-normal distribution shares the same various properties, like, for instance, closure under marginalization, conditioning and linear transformations \citep{gupta2000matrix}. The separability condition of the covariance matrix has two advantages. First, it allows the modeling of the temporal pattern of interest directly on the covariance matrix $\Phi$. Second, it represents a more parsimonious solution than that of the unrestricted $\Phi \otimes \Sigma$. Indeed, for that case the number of independent elements to compute would be $JT (JT+1)/2$, against $J(J+1)/2 + T(T+1)/2$ for the matrix-variate one. For example, setting $J=T=5$, one would have to estimate 325 elements in the multivariate case against 30 elements in the matrix-variate one.\\

Introduced by \cite{Viroli2011Oct}, the pdf of the finite Mixture of Matrix-Normals (MMN) model is given by
$$f(Z|\boldsymbol{\pi},\boldsymbol{\Theta})=\sum_{k=1}^K \pi_k \phi^{(J \times T)}(Z|M_k,\Phi_k,\Sigma_k),$$
where $\phi^{(J \times T)}$ represents the density function of a $J \times T$-dimensional matrix-variate normal, $K$ is the number of mixture components, $\boldsymbol{\pi}=\{\pi_k\}_{k=1}^{K}$ is the vector of mixing proportions, subject to constraint $\sum_{k=1}^{K}\pi_k=1$ and $\boldsymbol{\Theta}=\{\Theta_k\}_{k=1}^K$ is the set of component-specific parameters with $\Theta_k=\{M_k,\Phi_k,\Sigma_k\}$.

\subsection{The Mixture of Ordinal Matrices model}
Let denote by $y_{ijt}$ the observation of the $j$-th variable for the $i$-th unit at time $t$ ($i=1,\ldots,N$; $j=1,..,J$ and $t=1,\ldots,T$), that is: imagine to observe $N$ units and measuring $J$ different ordinal variables $T$ times throughout the course of the study. Let us reorganize this data in a random-matrix form such that $\mathbf{Y}=\{Y_i\}_{i=1}^{N}$ is a sample of $J\times T$-variate matrix observations $Y_i = (y_{ijt}) \in \mathbb{N}^{J\times T}$. The ordered classes are coded by non-negative integers such that each ordinal variable $J$ the ordinal levels are $\{1,2,\ldots,C_j\}$.\\
Then, we can assume that each variable $y_{ijt}$ is the manifestation of an underlying latent continuous variable $z_{ijt}$ which follows a Gaussian distribution, as done in the clustMD model \citep{clustmd}. At this point, we can assume that each observed ordinal matrix $Y_i$ is indeed the manifestation of a latent continuous random matrix $Z_i$, which follows a matrix-normal distribution.

\begin{equation*}
\mathbb{N}^{J\times T} \ni Y_i = 
 \begin{pmatrix}
  y_{i,1,1} & \cdots & y_{i,1,t} & \cdots & y_{i,1,T} \\
  \vdots  & \ddots & \vdots  & \cdots & \vdots  \\
  y_{i,j,1} & \cdots & y_{i,j,t} & \cdots & y_{i,j,T} \\
  \vdots  & \cdots & \vdots  & \ddots & \vdots  \\
  y_{i,J,1} & \cdots & y_{i,J,t} & \cdots & y_{i,J,T} 
 \end{pmatrix}
\longleftarrow 
Z_i = 
 \begin{pmatrix}
  z_{i,1,1} & \cdots & z_{i,1,t} & \cdots & z_{i,1,T} \\
  \vdots  & \ddots & \vdots  & \cdots & \vdots  \\
  z_{i,j,1} & \cdots & z_{i,j,t} & \cdots & z_{i,j,T} \\
  \vdots  & \cdots & \vdots  & \ddots & \vdots  \\
  z_{i,J,1} & \cdots & z_{i,J,t} & \cdots & z_{i,J,T} 
 \end{pmatrix} \in \mathbb{R}^{J\times T}
\end{equation*}
\vspace{.2cm}

To map from $Y_i$ to $Z_i$, let $\gamma_{j}$ denote a $C_{j} + 1$ -dimensional vector of thresholds that partition the real line for the $j$-th ordinal variable that has $C_j$ levels and let the threshold parameters be constrained such that $-\infty = \gamma_{j,0} \leq \gamma_{j,1} \leq\ldots\leq \gamma_{j,C_j}=\infty$. If the latent $z_{ijt}$ is such that $\gamma_{j,c-1}<z_{ijt}<\gamma_{j,c}$ then the observed ordinal response, $y_{ijt} = c$.

So, by assuming that each $Z_i$ follows a matrix-normal distribution, we can then cluster our data by means of finite Mixture of Matrix-Normals. 
In addition to $Z_i$, we introduce a latent binary K-dimensional vector that indicate whether the unit $i$ belongs to the $k$-th cluster, $\ell_i=(\ell_{i1},\ldots,\ell_{iK})$, such that $\ell_{ik}=1$ if the $i$-th unit belongs to the $k$-th cluster.\\

Moreover, let define $\mathcal{O}^{J \times T}$ the set of all possible  ordinal matrices of size $J\times T$ whose general row $j$ takes values in $\{1,\ldots,C_j\}$. 
Each element of $\mathcal{O}^{J \times T}$ is called a response pattern, that is each element of the set represents one of the possible configuration (pattern) of the $J\times T$ ordinal matrix, given the levels $C_j$. Let $R$ be the cardinality of $\mathcal{O}^{J \times T}$.
Each response pattern $Y_r \in \mathcal{O}^{J \times T}$ is generated by a portion $\Omega_r$ of the latent space $\mathbb{R}^{J \times T}$ according to thresholds $\boldsymbol{\gamma}:=\{\gamma_j \}_{j=1}^J$.
Let the binary vector $\tilde{Y}_i=(\tilde{Y}_{i1},\ldots,\tilde{Y}_{iR})$ 
be one-hot encoding of $Y_i$ such that if the $r$-th pattern is observed then $\tilde{Y}_{ir}=1$ and any other entry in the vector equals zero.
We can derive the joint density of $Z_i,\tilde{Y}_i,\ell_i$ as:
$$f(\tilde{Y}_i,Z_i,\ell_i)=f(\tilde{Y}_i|Z_i,\ell_i)f(Z_i|\ell_i)f(\ell_i).$$ 
Assuming that:
\begin{align*}
    &\ell_i \sim \mathcal{M}(1,\boldsymbol{\pi}), \, \boldsymbol{\pi}:=(\pi_1,\ldots,\pi_K)\\
    &Z_i|\ell_{ik}=1 \sim \mathcal{MN}_{(J \times T)}(Z_i|\Theta_k), \, \Theta_k :=\{M_k,\Phi_k,\Sigma_k\},\\
    &\tilde{Y}_i|Z_i,\ell_{ik}=1 \sim \mathcal{M}(1,\xi_i), \, \xi_i := (\mathbf{1}_{\Omega_1}(Z_i),...,\mathbf{1}_{\Omega_R}(Z_i)),
\end{align*}
we get:  
\begin{align*}
&f(\ell_i) =\prod_{k=1}^K \pi_k^{\ell_{ik}} \,;\\
&f(Z_i|\ell_{i}) = \prod_{k=1}^K\left[\phi^{(J \times T)}(Z_i|\Theta_k) \right]^{\ell_{ik}} \,;\\ 
&f(\tilde{Y}_{i}|Z_i,\ell_i) =\prod_{r=1}^R \mathbf{1}_{\Omega_r}(Z_i)^{\tilde{Y}_{ir}} \,,
\end{align*}

where $\mathcal{M}$ indicate the multinomial distribution and $\mathbf{1}_{\Omega_r}(Z_i)$ is the indicator function that equals 1 when the elements in $Z_i$ have values that determine the $r$-th pattern. Hence, when $\tilde{Y}_{ir} = 1$, the vector $\xi_i$ is a vector whose $r$-th element equals 1 and all the others equal 0.
In the following, $\boldsymbol{Z} := \{Z_i\}_{i=1}^N , \boldsymbol{\ell} := \{\ell_i\}_{i=1}^N$ and $\boldsymbol{\Theta} := \{\Theta_k,\pi_k\}_{k=1}^K$ will indicate the ensembles of $Z_i,\ell_i$ and of the parameters, respectively. Finally, let $\mathbf{\tilde{Y}} := \{\tilde{Y}_i\}^N_{i = 1}$ be the collection of the observed response pattern vectors $Y_i$.

\section{Inference}
\label{sec:inference}

\subsection{Thresholds}
\label{sec:thresholds}
\subsubsection{Identifiability}
A key point is of course the choice of the thresholds $\boldsymbol{\gamma}$.
Imagine to observe a sample of ordinal categories $c = {1,\dots,C_j}$ for variable $j$ and to work in the same framework as Section \ref{sec:model}. Let consider each variable separately in an univariate case for the sake of simplicity. Then, assume that each observation derive from the discretization of an underlying continuous variable following a normal distribution with parameters $(\mu_j, \sigma_j^2)$, and consider the $C_j - 1$ dimensional thresholds vector $\gamma_j$ as parameters to estimate together with the ones of the ones of the underlying normal. Then, the parameters set would be $\theta = (\mu_j, \sigma_j^2,\gamma_j)$, the parameter space $\Theta = (\mathbb{R},\mathbb{R}^+,\mathbb{R})$, and our model $P = \{p_{\theta} ; \theta \in \Theta\}$, with $p_{\theta}(y = c) = p(\gamma_{j,c-1} \leq z \leq \gamma_{,jc}), z \sim N(\mu_j,\sigma^2_j)$. It is clear that such a model would not be identifiable as there is no bijecton $\theta \mapsto p_{\theta}$. For instance, for a number of ordinal categories $C_j = 2$, $\theta_1 = (1.5, 1, 1.5)$ and $\theta_2 = (0,1,0)$ would yield the same distribution ($p_{\theta_1} = p_{\theta_2}$).\\

This simple example shows that we cannot aim at estimating the thresholds and the latent distribution parameters at the same time without incurring in some identifiability issues. Different strategies come to mind to overcome this problem.\\
Indeed, one solutions is to fix either the thresholds or the parameters $\boldsymbol{\Theta}$. In our case, being clearly the parameters of the mixture the quantity of interest, we decided to fix the thresholds as outlined in Section \ref{sec:thresholds}. However, it is also possible to go for a ``mixed strategy", partially fixing some of the distribution parameters and of the thresholds, to then estimate the rest, as done in as done in \cite{Millsap2004}.
\subsubsection{Choice of thresholds}
\label{sec:thresholds}
As written in Section \ref{sec:relworks}, assuming underlying continuous variables categorized according to some thresholds is not new and there are several ways of specifying such thresholds.\\
In \cite{clustmd} the thresholds $\boldsymbol{\gamma}=\{\gamma_j \}_{j=1}^J$ are fixed relying on data, by setting them as $\gamma_{j,c}=\varphi^{-1}(\delta_{j,c})$, where  $\delta_{j,c}$ is the proportion of variable $j$ which is less than or equal to level $c$ and $\varphi$ is the standard normal cumulative distribution function. With this assumption, the ordinal distribution of clusters will have the same global shape, not necessarily uni-modal, which makes clusters interpretation harder. \\

On the other hand, in \cite{bouveyron2020ordinal} thresholds are fixed arbitrarily (keeping equidistant the classes) as 
 $\gamma_j = (1.5,2.5,\dots ,C_j - 0.5)$ and $C_j$ is assumed to be equal for all variables, proposing a scale conversion pre-processing algorithm (\cite{gilula2019study}) for cases when this does not hold true. The advantages of such an approach is that an underlying space is related with the range of the ordinal entries, leading to easily interpretable results. Another result of equidistant thresholds is that it produces monotonicity around the mode, creating more separated and interpretable clusters. In the following work this approach will be followed.\\
 It is important to remark that this choice of thresholds does not impose any constraint on the distribution of the ordinal levels, but the monotonic behaviour around the mode.\\
 Finally, it is also worth noting that the thresholds are fixed and do not change over time.
 
\subsection{EM-algorithm}
\label{sec:em}
The EM algorithm (\cite{EM}) is an iterative algorithm alternates two steps: the expectation step (E-step) and the maximization step (M-step).
It start from an initialization $\hat{\boldsymbol{\Theta}}^{(0)}$ of the parameters. Then, let denote with the superscript ${(s + 1)}$ the parameters estimated in the current step and with $(s)$ the ones computed in the previous step.\\
The E-step consists of evaluating $\mathcal{Q}(\boldsymbol{\Theta},\hat{\boldsymbol{\Theta}}^{(s)}) := \mathbb{E}( \log \mathcal{L}_C(\boldsymbol{\Theta};\mathbf{\tilde{Y}},\mathbf{Z},\boldsymbol{\ell})|\hat{\boldsymbol{\Theta}}^{(s)},\mathbf{\tilde{Y}})$, that is the expectation of the complete log-likelihood 
%with respect to the latent data $Z$ and the cluster labels $\ell$, 
conditioned on the parameters computed in the previous step and on the observed data. In the M-step the parameters are updated by maximizing the expected log-likelihood found on the E step, that is $\hat{\boldsymbol{\Theta}}^{(s+1)} := \underset{\Theta}{\arg \max} \, \mathcal{Q}(\boldsymbol{\Theta},\hat{\boldsymbol{\Theta}}^{(s)})$.\\The iteration process is repeated until convergence on the log-likelihood is met.

\subsection{Complete Likelihood}
\label{sec:lik}

The complete log-likelihood can be written as 
\begin{multline*}
\log\mathcal{L}_C(\boldsymbol{\Theta};\mathbf{\tilde{Y}},\mathbf{Z},\boldsymbol{\ell}) = \sum_{i=1}^N \Biggl\{ \sum_{r=1}^R \tilde{Y}_{ir} \mathbf{1}_{\Omega_r}(Z_i) 
+ \sum_{k=1}^{K} \ell_{ik} \Biggl[ \log(\pi_k) - \frac{TJ}{2}\log(2\pi)-\frac{J}{2}\log(|\Phi_k|) - \\
\frac{T}{2}\log(|\Sigma_k|)
-\frac{1}{2} tr[\Sigma^{-1}_k(Z_i-M_k)\Phi^{-1}_k(Z_i-M_k)^{\intercal}] \Biggl] \Biggl\}.
\end{multline*}

\subsection{E-step computation}
Conditioning on the parameters computed in the step $(s)$, at the step $(s+1)$ the value of $\mathcal{Q}(\boldsymbol{\Theta},\boldsymbol{\Theta}^{(s)})$ is:

\begin{align}
& \mathcal{Q}(\boldsymbol{\Theta},\boldsymbol{\Theta}^{(s)}) := \mathbb{E}( \log \mathcal{L}_C(\boldsymbol{\Theta};\mathbf{\tilde{Y}},\mathbf{Z},\boldsymbol{\ell})|\hat{\boldsymbol{\Theta}}^{(s)},\mathbf{\tilde{Y}}) = \nonumber \\
&\mathbb{E}\Biggl(\sum_{i=1}^N \Biggl\{ \sum_{r=1}^R \tilde{Y}_{ir} \mathbf{1}_{\Omega_r}(Z_i)+ \sum_{k=1}^{K} \ell_{ik} \Biggl[ \log(\hat{\pi}^{(s)}_k) - \frac{TJ}{2}\log(2\pi)-\frac{J}{2}\log(|\hat{\Phi}^{(s)}_k|) - \nonumber\\ 
& \hspace{.5cm} \frac{T}{2}\log(|\hat{\Sigma}^{(s)}_k|) -\frac{1}{2} \text{tr}[\hat{\Sigma}^{-1(s)}_k(Z_i-\hat{M}^{(s)}_k) \times \hat{\Phi}^{-1(s)}_k(Z_i-\hat{M}^{(s)}_k)^{\intercal}] \Biggl] \Biggl\} \Biggl| \hat{\boldsymbol{\Theta}}^{(s)},\mathbf{\tilde{Y}}\Biggl) = \label{eq:eloglik} \\
& \sum_{i=1}^N \sum_{r=1}^R \tilde{Y}_{ir} \, \mathbb{E}(\mathbf{1}_{\Omega_r}(Z_i)|\hat{\boldsymbol{\pi}}^{(s)},\hat{\boldsymbol{\Theta}}^{(s)},\mathbf{\tilde{Y}}) \, + \label{e1} \\ 
&  
\sum_{i=1}^N \sum_{k=1}^{K} \mathbb{E}(\ell_{ik}|\hat{\boldsymbol{\pi}}^{(s)},\hat{\boldsymbol{\Theta}}^{(s)},\mathbf{\tilde{Y}}) \,\times \nonumber \\
& \hspace{1.5cm} \Biggl[ \log(\hat{\pi}^{(s)}_k) - \frac{TJ}{2} \times \log(2\pi)-\frac{J}{2}\log(|\hat{\Phi}^{(s)}_k|)- \frac{T}{2}\log(|\hat{\Sigma}^{(s)}_k|)\Biggl] \, -\label{e2} \\
& \sum_{i=1}^N \sum_{k=1}^{K} \frac{1}{2}\, \mathbb{E}(\ell_{ik}\text{tr}[\hat{\Sigma}^{-1(s)}_k(Z_i-\hat{M}^{(s)}_k)  \times \hat{\Phi}^{-1(s)}_k(Z_i-\hat{M}^{(s)}_k)^{\intercal}] |\hat{\boldsymbol{\Theta}}^{(s)},\mathbf{\tilde{Y}}) \label{e3}  
\end{align}

We can treat each of the three expectations separately, and we get for (\ref{e1})
\begin{align*}
  \mathbb{E}(\mathbf{1}_{\Omega_r}(Z_i)|\hat{\boldsymbol{\Theta}}^{(s)},\mathbf{\tilde{Y}}) %&= 1 \cdot \mathbb{P}(Z_i \in \Omega_r|Y^R_i,\dots) + 0 \cdot \mathbb{P}(Z_i \notin \Omega_r|Y^R_i,\dots) = \\
  & = \mathbb{P}(Z_i \in \Omega_r|\hat{\boldsymbol{\Theta}}^{(s)},\tilde{Y}_i).
  %= \sum_{k=1}^K \pi_k \int_{\Omega_r} f(Z|\Theta_k)dZ
\end{align*}
Since we are conditioning on $\tilde{Y}_i$, the observed response pattern is known and therefore the probability of $Z_i$ belonging to $\Omega_r$ is equal to 1 when $\tilde{Y}_{ir} = 1$ and 0 otherwise.\\

For (\ref{e2}), we can write
\begin{align}
\begin{split}
    \mathbb{E}(\ell_{ik}&|\tilde{Y}_{ir}=1,\hat{\boldsymbol{\Theta}}^{(s)}) = \mathbb{P}(\ell_{ik} = 1|\tilde{Y}_{ir}=1,\hat{\boldsymbol{\Theta}}^{(s)}) \\[10pt]
    &= \frac{\mathbb{P}(\ell_{ik} = 1|\hat{\boldsymbol{\Theta}}^{(s)}) \mathbb{P}(Y^R_{ir} =  1|\ell_{ik}  1,\hat{\boldsymbol{\Theta}}^{(s)})}{\mathbb{P}(Y^R_{ir} = 1|\hat{\boldsymbol{\Theta}}^{(s)})}  \\[10pt]
    & = \frac{\pi_k^{(s)} \int_{\Omega_r} f(Z|\Theta^{(s)}_k)dZ}{\sum_{k=1}^K \pi^{(s)}_k \int_{\Omega_r} f(Z|\Theta^{(s)}_k)dZ} =: {\uptau}^{(s+1)}_{ik},
\end{split}
 \end{align}
where the integral can be approximated through a Monte-Carlo approach applied on the vectorized reparametrization of the matrix-variate distribution.\\

On the other hand, (\ref{e3}) is less straightforward, and we will need some tricks to deal with it. As done in \cite{clustmd}, we can break down as
\begin{align}
\mathbb{P}&(\ell_{ik}=1  |\hat{\boldsymbol{\Theta}}^{(s)},\mathbf{\tilde{Y}}) \, \times \nonumber \\ 
& \mathbb{E}(tr[\hat{\Sigma}^{-1(s)}_k(Z_i-\hat{M}^{(s)}_k) \times \hat{\Phi}^{-1(s)}_k(Z_i-\hat{M}^{(s)}_k)^{\intercal}] |\ell_{ik}=1,\mathbf{\tilde{Y}},\hat{\boldsymbol{\Theta}}^{(s)}). 
\end{align}

By opening the matrix product in the second term we get:\\
\begin{align}
& \hat{\Sigma}^{-1(s)}_k(Z_i-\hat{M}^{(s)}_k)\hat{\Phi}^{-1(s)}_k(Z_i-\hat{M}^{(s)}_k)^{\intercal} = \nonumber \\
& \; \, \hat{\Sigma}^{-1(s)}_k Z_i \hat{\Phi}^{-1(s)}_k Z^{\intercal}_i - \hat{\Sigma}^{-1(s)}_k  Z_i \hat{\Phi}^{-1(s)}_k \hat{M}^{\intercal}_k  
-\hat{\Sigma}^{-1(s)}_k \hat{M}^{(s)}_k \hat{\Phi}^{-1(s)}_k Z^{\intercal}_i + \hat{\Sigma}^{-1(s)}_k \hat{M}_k 
\hat{\Phi}^{-1(s)}_k
\hat{M}^{\intercal}_k.
\end{align}

It is easy to realize that its solution requires the computation of $\mathbb{E}(Z_i|\ell_{ik} = 1,\hat{\boldsymbol{\Theta}}^{(s)},\tilde{Y}_{ir} = 1)$ and of the expectation of a matrix quadratic forms, specifically for $\mathbb{E}(Z_i\hat{\Phi}^{-1(s)}_k Z_i^{\intercal}|\ell_{ik} = 1,\hat{\boldsymbol{\Theta}}^{(s)},\tilde{Y}_{ir} = 1)$. As we will in Section \ref{sec:mstep}, we will also need to compute $\mathbb{E}( Z_i^{\intercal}\hat{\Sigma}^{-1(s+1)}_kZ_i|\ell_{ik} = 1,\hat{\boldsymbol{\Theta}}^{(s)},\tilde{Y}_{ir} = 1)$ for the M-step. The computation of the expectation of $Z_i$ and of such quadratic form necessitates in turn to compute the moments of a truncated matrix-variate Gaussian. However, that is a complex task, so we will need to work the issue around.\\

We can bypass the problem concerning the expectation of $Z_i$ by defining with $z_i \in \mathbb{R}^{JT \times 1}$ the vectorized version of $Z_i$ and computing
\begin{equation}
    \mathbb{E}(z_i|\ell_{ik} = 1,\tilde{Y}_{ir} = 1,\hat{\boldsymbol{\Theta}}^{(s)}) =: m^{(s+1)}_{ik}
\end{equation}
through the use of a Monte Carlo approach and specifically the use of a Gibbs sampler to sample from a truncated multivariate normal distribution. Moreover, the samples generated to calculate the first moment $m_{ik}^{(s+1)}$ can be reused to compute the matrix $S_{ik}^{(s+1)} := \mathbb{E}(z_i z_i^{\intercal}|\ell_{ik} = 1,\tilde{Y}_{ir} = 1,\hat{\boldsymbol{\Theta}}^{(s)}) \in \mathbb{R}^{JT \times JT}$, that can be approximated by calculating the inner product of the vectors used to compute $m^{(s+1)}_{ik}$ then calculating the sample mean of these inner products.\\

Subsequently, we can find $\mathbb{E}(Z_i\hat{\Phi}^{-1(s)}_k Z_i^{\intercal}|\ell_{ik} = 1,\hat{\boldsymbol{\Theta}}^{(s)},\tilde{Y}_{ir} = 1)$ by computing it element-by-element. In order to do that, we can define 
 $D^{(s+1)}_{ik} := \mathbb{E}(Z_i\hat{\Phi}^{-1(s)}_k Z_i^{\intercal}|\ell_{ik} = 1,\hat{\boldsymbol{\Theta}}^{(s)},\tilde{Y}_{ir} = 1))$, $\hat{\varphi}^{(s)}_{k,gd}$ as the $(g,d)^{th}$ element of  $\hat{\Phi}^{-1(s)}_k$. Then, the $(h,t)^{th}$ element of $Z_i^{\intercal}\hat{\Phi}^{-1(s)}_k Z_i$ would be $\sum_{d=1}^T \sum_{g=1}^T z_{i,hg} \hat{\varphi}^{(s)}_{k,gd} z_{i,td}$ and we would get

\begin{align}
&D^{(s+1)}_{ik} := \mathbb{E}(Z_i\hat{\Phi}^{-1(s)}_k Z_i^{\intercal}|\ell_{ik} = 1,\hat{\boldsymbol{\Theta}}^{(s)},\tilde{Y}_{ir} = 1)) \nonumber \\
& = \mathbb{E}\Biggl(\Big(\sum_{d=1}^T \sum_{g=1}^T z_{i,hg} \hat{\varphi}^{(s)}_{k,gd} z_{i,td} \Big)_{h,t} |\ell_{ik} = 1,\hat{\boldsymbol{\Theta}}^{(s)},\tilde{Y}_{ir} = 1 \Biggl) \nonumber \\
& = \mathbb{E}\Biggl(\Big(\sum_{d=1}^T \sum_{g=1}^T z_{i,hg} z_{i,td} \hat{\varphi}^{(s)}_{k,gd} \Big)_{h,t} |\ell_{ik} = 1,\hat{\boldsymbol{\Theta}}^{(s)},\tilde{Y}_{ir} = 1\Biggl) \nonumber \\ 
& = \Big(\sum_{d=1}^T \sum_{g=1}^T S^{(s+1)}_{ik,[(g-1)J+h,(d-1)J+t]}\hat{\varphi}^{(s)}_{k,gd} \Big)_{h,t}, \label{eq:dsimple}
\end{align}

where in we make use of the the elements of $S_{ik}$.\\

As written above, we would also need to compute $\mathbb{E}( Z_i^{\intercal}\hat{\Sigma}^{-1(s+1)}_kZ_i|\ell_{ik} = 1,\hat{\boldsymbol{\Theta}}^{(s)},\tilde{Y}_{ir} = 1)$, which we can do by following the same reasoning. By defining $C_{ik}^{(s+1)} := \mathbb{E}(Z_i^{\intercal}\hat{\Sigma}^{-1(s+1)}_k Z_i|\ell_{ik} = 1,\hat{\boldsymbol{\Theta}}^{(s)},\tilde{Y}_{ir} = 1)$ and by denoting by $\hat{\sigma}^{(s+1)}_{k,gd}$ the $(g,d)^{th}$ element of $\hat{\Sigma}^{-1(s+1)}_k$. Then, the $(h,t)^{th}$ element of $Z_i^{\intercal}\hat{\Sigma}^{-1(s)}_k Z_i$ is $\sum_{d=1}^J \sum_{g=1}^J z_{i,gh} \hat{\sigma}^{(s+1)}_{k,gd} z_{i,dt}$, and we get

\begin{align}
&C_{ik}^{(s+1)}:=  \mathbb{E}(Z_i^{\intercal}\hat{\Sigma}^{-1(s+1)}_k Z_i|\ell_{ik} = 1,\hat{\boldsymbol{\Theta}}^{(s)},\tilde{Y}_{ir} = 1) \nonumber \\
& = \mathbb{E}\Biggl(\Big(\sum_{d=1}^J \sum_{g=1}^J z_{i,gh} \hat{\sigma}^{(s+1)}_{k,gd} z_{i,dt} \Big)_{h,t} |\ell_{ik} = 1,\hat{\boldsymbol{\Theta}}^{(s)},\tilde{Y}_{ir} = 1 \Biggl) \nonumber \\
& = \mathbb{E}\Biggl(\Big(\sum_{d=1}^J \sum_{g=1}^J z_{i,gh} z_{i,dt} \hat{\sigma}^{(s+1)}_{k,gd}\Big)_{h,t}|\ell_{ik} = 1,\hat{\boldsymbol{\Theta}}^{(s)},\tilde{Y}_{ir} = 1\Biggl) \nonumber \\
& = \Big(\sum_{d=1}^T \sum_{g=1}^T S^{(s+1)}_{ik,[(h-1)J+g,(t-1)J+d]}\hat{\sigma}^{(s+1)}_{k,gd} \Big)_{h,t}.  \label{eq:csimple}
\end{align}

Finally, this means that computing $\mathcal{Q}(\boldsymbol{\Theta},\hat{\boldsymbol{\Theta}}^{(s)})$ requires to compute:
\begin{itemize}
\item $\mathbb{E}(\ell_{ik}|\tilde{Y}_{ir} = 1,\hat{\boldsymbol{\Theta}}^{(s)}) %\frac{\pi_k^{(s)} \int_{\Omega_r} f(Z|\Theta^{(s)}_k)dZ}{\sum_{k=1}^K \pi^{(s)}_k \int_{\Omega_r} f(Z|\Theta^{(s)}_k)dZ} 
=: {\uptau}^{(s+1)}_{ik}$,
\item $\mathbb{E}(z_i|\ell_{ik} = 1,\tilde{Y}_{ir} = 1,\hat{\boldsymbol{\Theta}}^{(s)}) =: m^{(s+1)}_{ik}$,
\item $\mathbb{E}(z_i z_i^{\intercal}|\ell_{ik} = 1,\tilde{Y}_{ir} = 1,\hat{\boldsymbol{\Theta}}^{(s)}) =: S_{ik}^{(s+1)}$, whose elements are required for the computation of $D^{(s+1)}_{ik}$ and $C^{(s+1)}_{ik}$. %that can be calculated by subtracting the second moment to the first one
\end{itemize}

\subsection{M-step}
\label{sec:mstep}
By taking the first derivatives of Equation \ref{eq:eloglik}, the maximum likelihood estimators of the parameters are given by

\begin{align}
&\hat{\pi}^{(s+1)}_k = \frac{\sum_{i=1}^N  \hat{\uptau}^{(s+1)}_{ik}}{N}
\label{eq:pi} \\
&\hat{M}^{(s+1)}_k = \frac{\sum_{i=1}^N \hat{\uptau}^{(s+1)}_{ik} \hat{M}^{(s+1)}_{ik}}{\sum_{i=1}^N \hat{\uptau}^{(s+1)}_{ik}}
\end{align}
\label{eq:m}

where $\hat{M}^{(s+1)}_{ik} := \mathbb{E}(Z_i|\ell_{ik} = 1,\tilde{Y}_{ir} =1,\hat{\boldsymbol{\Theta}}^{(s)}) =  \mathbb{E}(\text{vec}_{J \times T}^{-1}(z_i)|\ell_{ik} = 1,\tilde{Y}_{ir} =1,\hat{\boldsymbol{\Theta}}^{(s)}) = \text{vec}_{J \times T}^{-1}(m_{ik})$, and $\text{vec}_{J \times T}^{-1}$ is the inverse of the vectorization function, i.e. the function mapping from a $JT$-dimensional vector to a $J \times T$ matrix. The two covariance matrices are interdependent and require the computation of $C_{ik}^{(s+1)}$ and $D_{ik}^{(s+1)}$. The updating of the covariance matrices is obtained through: 

\begin{align}
&\frac{\hat{\Sigma}^{(s+1)}_k= \sum_{i=1}^N \uptau^{(s+1)}_{ik}[D_{ik}^{(s+1)}- \hat{M}^{(s+1)}_k \hat{\Phi}^{-1(s)}_k M^{\intercal(s+1)}_{ik} - M^{(s+1)}_{ik} \hat{\Phi}^{-1(s)}_k \hat{M}^{\intercal(s+1)}_k +\hat{M}^{(s+1)}_k \hat{\Phi}^{-1(s)}_k \hat{M}^{\intercal(s+1)}_k]}{T\sum_{i=1}^N \uptau^{(s+1)}_{ik}}, \label{eq:sigma_update}\\
&\frac{\hat{\Phi}^{(s+1)}_k= \sum_{i=1}^N \uptau^{(s+1)}_{ik}[C_{ik}^{(s+1)} - \hat{M}^{\intercal(s+1)}_k \hat{\Sigma}^{-1(s+1)}_k M^{(s+1)}_{ik} - M^{\intercal(s+1)}_{ik} \Sigma^{-1(s+1)}_k \hat{M}^{(s+1)}_k + \hat{M}^{\intercal(s+1)}_k \hat{\Sigma}^{-1(s+1)}_k \hat{M}^{(s+1)}_k]}{J\sum_{i=1}^N \uptau^{(s+1)}_{ik}}. \label{eq:phi_update}
\end{align}

It is worth to remark that the computation of $\hat{\Sigma}^{(s+1)}_k$ and $\hat{\Phi}^{(s+1)}_k$ relies on $D^{(s+1)}_{ik}$ and $C^{(s+1)}_{ik}$, respectively. The two quantities in turn rely on the elements of $\hat{\Phi}^{(s)}_k$ and $\hat{\Sigma}^{(s+1)}_k$, as shown in Equation \ref{eq:dsimple} and Equation \ref{eq:csimple}. This means that in the algorithm one needs to compute first $D^{(s+1)}_{ik}$, then $\hat{\Sigma}^{(s+1)}_k$, and $C^{(s+1)}_{ik}$ and $\hat{\Phi}^{(s+1)}_k$ subsequently. The updating order of the parameters can be exchanged, but it is important to use the updated parameters coherently.

\subsection{Initialization}
\label{sec:init}
To find the initial values of $\hat{\boldsymbol{\Theta}}^{(0)}$ mentioned in Section \ref{sec:em}, our proposal is the following.
Identity matrices are chosen for the initialization of the covariance matrices $\Phi_k$ and $\Sigma_k$, while $\pi_k = 1/K$.
For the initialization of $M_k$, two solutions are proposed and tested in Section \ref{sec:sub_init}. The first is a Kmeans++ \citep{Arthur2007Jan} initialization, that is performed on the vectorized data. The second is a multiple random initialization: the mean matrices $M_k$ are chosen by uniform sampling $K$ matrices among the $N$ observed data matrices. Since the EM algorithm is not guaranteed to converge toward a global optimum, the algorithm is applied multiple times and the results with the highest log-likelihood is selected. For simulations in Section \ref{sec:sub_init}, 5 random initialization proved to be enough, but for more complex setting a higher number might be needed.

\subsection{Selection of the number of cluster K} \label{sec:choiceK}
The number of cluster $K$ is selected by minimizing the BIC \citep{Schwarz1978Mar} criterion. The BIC for a number of cluster $k$ is defined as
\begin{equation*}
    \text{BIC}_k := -2\log\mathcal{L}_{O}(\boldsymbol{\Theta;\boldsymbol{\tilde{Y}}}) + \nu_k \log(N),
\end{equation*}
where $\nu_k$ is the total number of model parameters: 
\begin{equation}
\label{eq:num_par}
    \nu_k :=  k[1 + JT + J(J + 1)/2 + T(T + 1)/2]-1,
\end{equation}
and $\mathcal{L}_{O}(\boldsymbol{\Theta;\boldsymbol{\tilde{Y}}})$ is the observed likelihood of the model, that is

\begin{equation*}
\mathcal{L}_{O}(\boldsymbol{\Theta;\boldsymbol{\tilde{Y}}}) := \prod_{i = 1}^N \prod_{r = 1}^R\Biggl(\sum_{k=1}^K \pi_k \int_{\Omega_r} f(Z|\Theta_k)dZ\Biggl)^{\tilde{Y}_{ir}}.
\end{equation*} 
%since $\tilde{Y}_i$ would still have multinomial distribution but, not conditioning on $Z_i$ and $\ell_{ik}$, $\xi_i$ is not deterministic.\\
To select the model with the optimal $K$, the algorithm needs to be executed for every $k = 1,...,K$ and the model with the lowest $\text{BIC}_k$ is chosen.

\subsection{Classification}
\label{sec:classification}
Finally, a criterion for the classification of the units must be established. The criterion we use is the maximum conditional allocation probability. Defining with the superscript $c$ the step at which the convergence has been reached or the maximum number of iterations attained, the observation $i$ will be allocated to the cluster $h = \arg\max \limits_{h} \, \uptau^{(c)}_{ih}$.

\section{Evaluation}
\label{sec:evaluation}

This section presents numerical experiments on simulated data in order to illustrate the
behavior of the proposed model regarding the influence of the initialization procedure and sample size, the robustness to different noise ratio in the data, the model selection and in comparison with its continuous counterpart when used on ordinal data treated like quantitative data.\\
The algorithm has been implemented in R.

\subsection{Simulation Setup}
\label{sec:sim_set}
100 different samples have been simulated for increasing number of units $N \in \{300,1500,3000 \}$, with $K = 3$, $J = 5$, $T = 5$, $\pi = (0.3,0.4,0.3)$ and $C_j = 5$ levels $\forall j=1,\ldots,J$ . Each sample has been drawn from a matrix-variate Gaussian and then discretized according to the thresholds chosen in Section \ref{sec:thresholds}.Concerning the distributions' parameters, identity matrices were chosen for matrices $\Phi_k$ and $\Sigma_k$ for every cluster, while the mean matrices $M_k$ were selected so that there would be a partial overlap among the clusters, in order to avoid triviality. However, estimating theoretically the overlapping area in such a setting is complex endeavour. That is why we evaluate an approximated ``optimal" Adjusted Rand Index (ARI) (\cite{Rand1971Dec}), by 
comparing the classification obtained using the true model parameters with the known groups. 
Thus, the mean matrices $M_k$ are chosen so that this estimated optimal ARI would be around 0.85. Note that we would expect the study to show
convergence to this number as the sample size increases. This setting led to the choice of $M_1 = 1.75 \cdot \mathbf{1}_5 \mathbf{1}_5^\top, M_2 = 2.5 \cdot \mathbf{1}_5 \mathbf{1}_5^\top$ and $M_3 = 3.25 \cdot \mathbf{1}_5 \mathbf{1}_5^\top$, where $\mathbf{1}_5$ is a 5-dimensional vector whose elements
are all 1.\\
Moreover, three scenarios are derived from this setting by adding  some noise fraction within the clusters by simulating a proportion $\tau$ of units using a uniform distribution on levels $C_j$, allocated to the three clusters proportionally to the clusters' size: 0 (scenario 1), 0.1 (scenario 2), 0.2 (scenario 3).\\
The two different kinds of initialization described in Section \ref{sec:init} have been tested.

Finally, we use a difference between observed log-likelihood at step $(s+1)$ and $(s)$ as stopping criterion, setting this difference to be lower than 0.001 as stopping rule.

Regarding the algorithm setup, we set to 100 iterations as the burn-in period of Gibbs sampler in the E-step, and a thinning equal to 2 to prevent too correlated samples. The number of simulated samples is set to 100. 
Computation time for one iteration on 2.40 GHz 11th Gen Intel Core i5-1135G7 with 16 Go RAM for one step of the algorithm with Kmeans++ initialization is about 8 seconds for $N  = 300$ and about 80 seconds for $N = 3000$. 

 \subsection{Influence of initialization \& sample size}
 \label{sec:sub_init}
\begin{figure}[ht!]
\centering
\includegraphics[width = 1\linewidth]{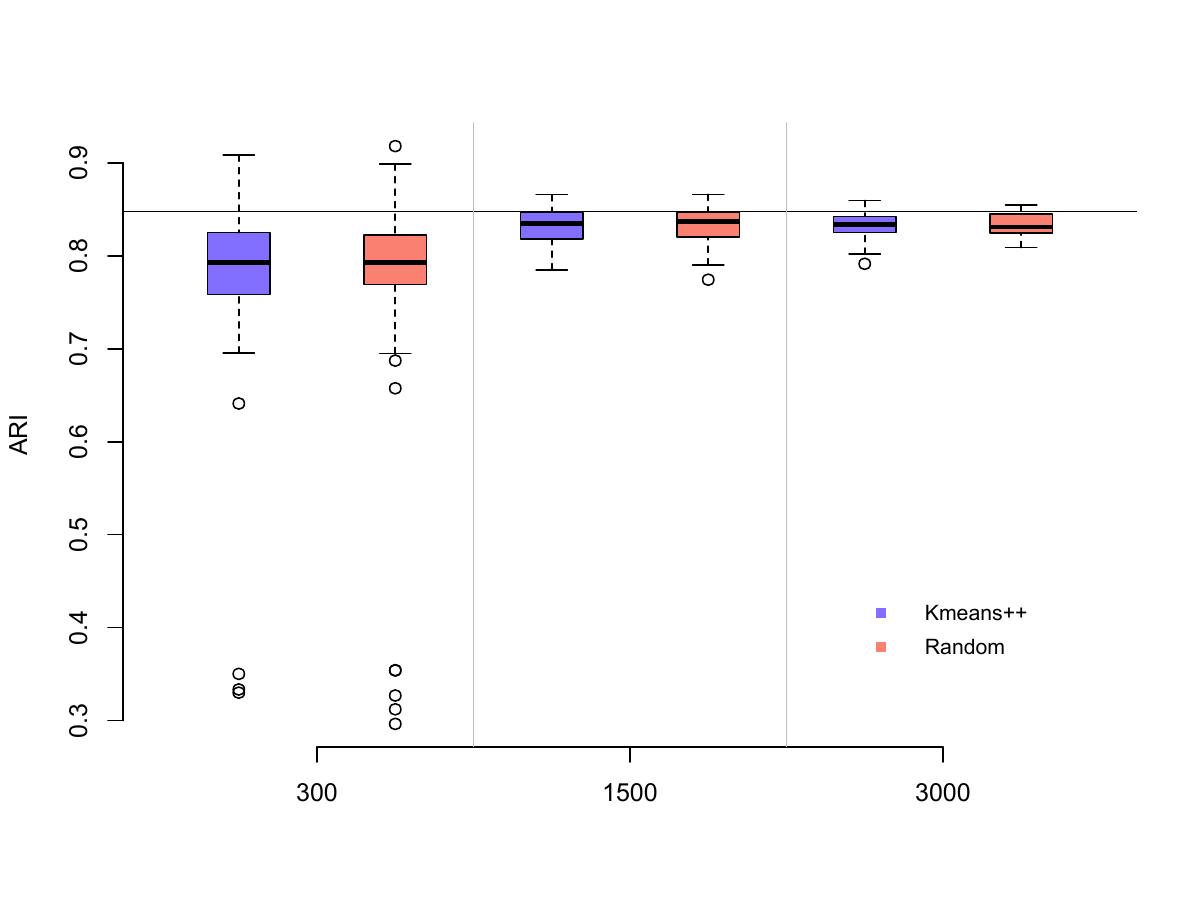}
\caption{Influence of initialization. The horizontal line represents the estimated optimal ARI.}
\label{fig:inf_init}
\end{figure}

 This first experiment aims at studying the ability of MOM to recover the simulated model depending on the type of initialization of the EM algorithm. Figure \ref{fig:inf_init} shows the quality of estimated partitions assessed by means of ARI. We recall that an ARI of 1 indicates that the partition provided by the algorithm is perfectly aligned with the simulated one. Conversely, an ARI of 0 indicates that the two partitions could as well be some random matches. On the graph, the optimal ARI ($\approx 0.85$) according to the simulation scheme is represented by a horizontal line. The boxplots do not seem to show any significant difference in the median values of the ARI measurements between the two initialization methods, but for sample size equal to 300 there seems to be a greater variability in the results, probably steaming from the smaller sample size.

Overall, from a partitioning point of view, the two initialization techniques do not seem to produce significantly different results. We decided to measure their performance also by computing the Mean Absolute Percentage Error (MAPE) on their estimation of the distribution parameters. The MAPE calculates the average percentage difference between the actual and predicted values of a variable, therefore providing a relative measure of error. For a sample of N units, for a generic parameter $\theta$ it is expressed through the formula:

\begin{equation*}
    \mbox{MAPE}={\frac {1}{N}}\sum _{i=1}^{N}\left|{\frac {\theta_{i}-\hat{\theta}_{i}}{\theta_{i}}}\right|,
\end{equation*}

where $\hat{\theta}_i$ is the estimated parameter %based on the values of the i-th unit
and $\theta_i$ is the true parameter. 
MAPE has some limitations, such as the fact that it cannot be used when actual values are zero or close to zero. This is why for the covariance matrices only the diagonal elements are considered.

\begin{figure*}[!ht]
\hspace*{-.2in}
\begin{subfigure}{.5\textwidth}
  \centering
  \includegraphics[width=1\linewidth]{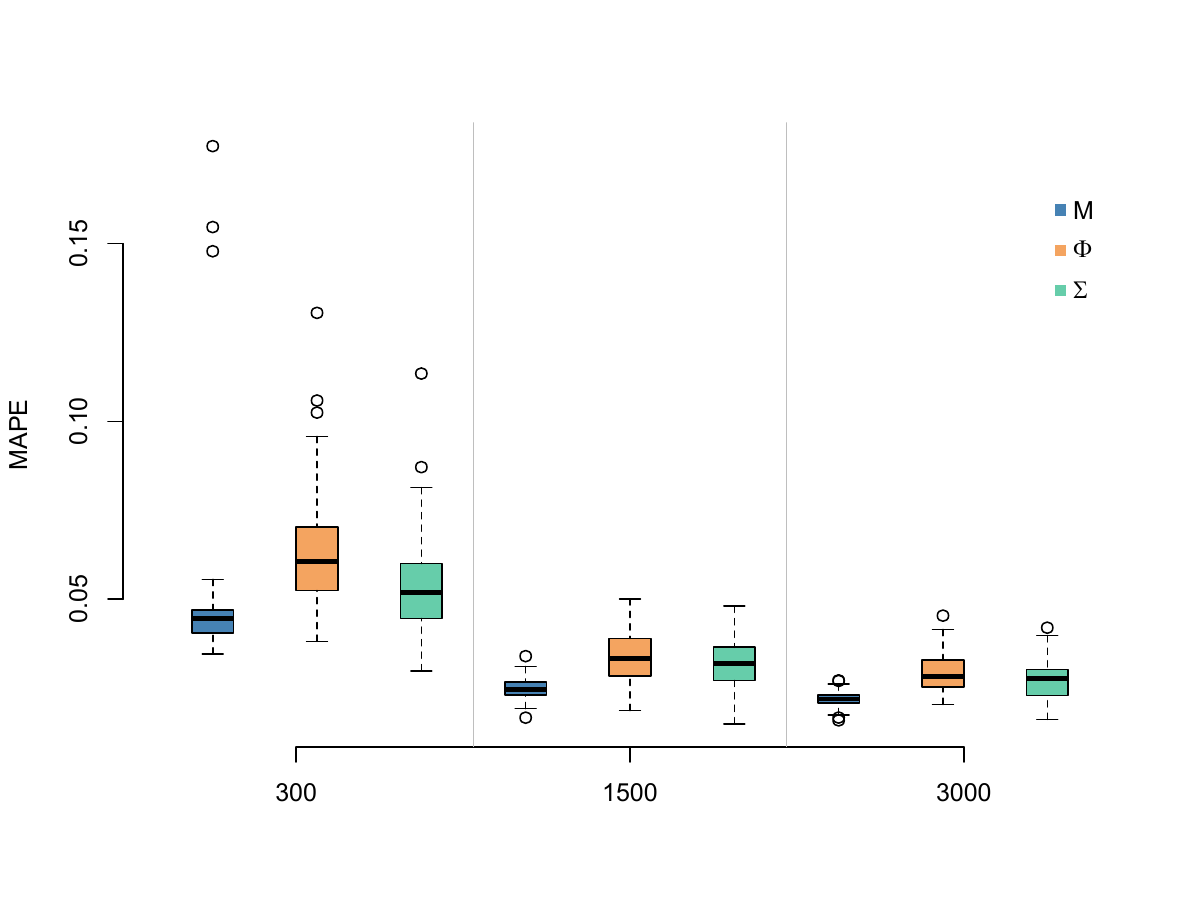}
  \caption{Kmeans++ init}
  \label{fig:MAPE_kmeans}
\end{subfigure}%
\hspace*{.3in}
\begin{subfigure}{.5\textwidth}
  \centering
  \includegraphics[width=1\linewidth]{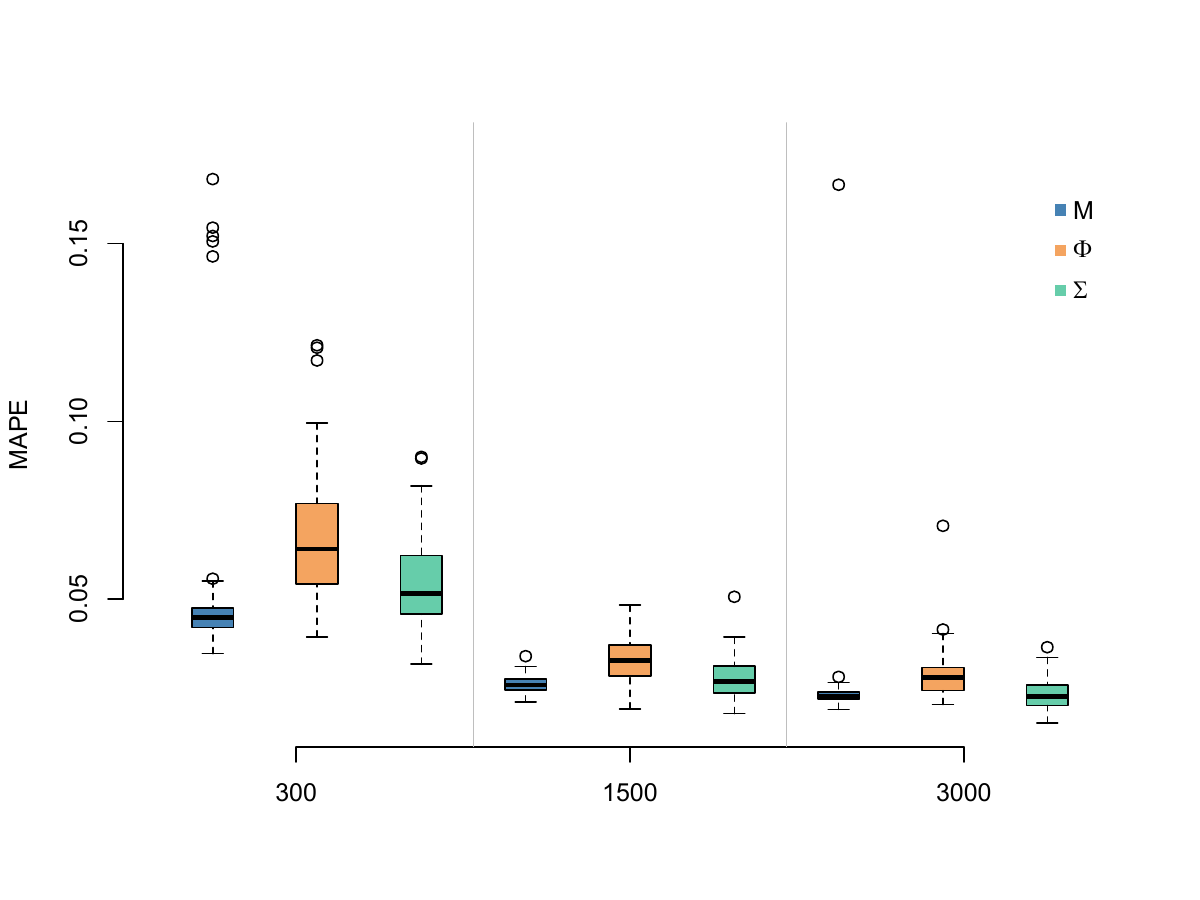}
  \caption{Random init.}
  \label{fig:MAPE_init}
\end{subfigure}
\hspace*{.5in}
\caption{MAPE for increasing N}
\label{fig:MAPE_N_init}
\end{figure*}

Results are shown in Figure \ref{fig:MAPE_N_init}. There seems not to be a clear difference between the two initializations.\\
Concerning the influence of the sample size, the model behaves as expected: as the sample size increases, the partitioning capabilities improve and tend towards the optimal error. The same happens when we observe the errors concerning the parameter estimations for both the initialization procedures.

Globally, there not seems to be a significant difference in terms of performance results for the two initialization procedures regarding the partitioning capabilities. The only biggest difference seems to be the slightly lower variability of the estimates produced by the random initialization. Nonetheless, it is worth noticing that the random initialization is to some extent a greedy procedure which requires to compute the algorithm several times with the purpose of selecting the best result, and therefore, depending on the number of random initializations chosen, it can easily become time-consuming and computationally costly. \\
In the following, given the similarities in performance and the computational advantages, we will carry out most of the analysis using only the Kmeans++ initialization.

\subsection{Robustness to noise}
\begin{figure}[ht!]
\centering
\captionsetup{justification=centering}
\includegraphics[width = 1\linewidth]{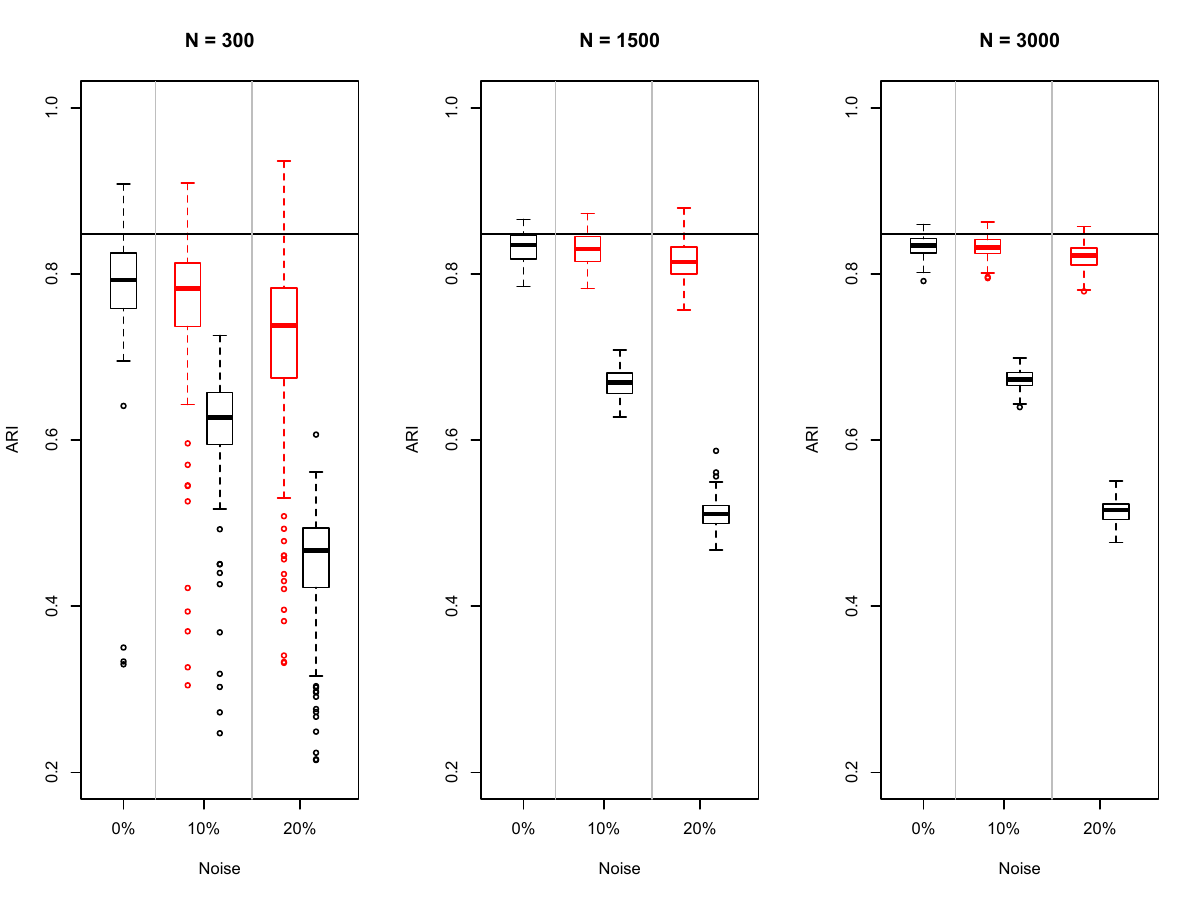}
\caption{ARI for increasing noise proportions and increasing N.\\
The red (left) box plots is for non-noisy units (0.1 and 0.2 of noise), the black (right) for all units.}
\label{fig:ARI_noise}
\end{figure}

As written in Section \ref{sec:sim_set}, we also simulated some noisy data to study the behaviour of MOM when the underlying normality assumption is not fully respected.
ARIs for different noise proportions were measured and the results are visible in Figure \ref{fig:ARI_noise}. We decided to measure two quantities: the overall ARI for all the units and the ARI just for the non-noisy ones.

As we would expect, the overall quality of partitioning estimates decreases as the level of noise increases, indicating that MOM is actually disturbed by the noise.

Interestingly, for N large enough, the model proofs itself robust and it classes perfectly non-noisy data, reaching the optimal ARI, represented by the horizontal black line in the graph. For $N = 300$, the noise disturbs the model estimate, and we do not get an ARI as close to the optimal one as for bigger samples, but still overall better for non-noisy data.
The clustering of matrix-normally distributed data therefore seems a bit disturbed by noise when $N$ is small, but it corrects when $N$ increases. This may be due to the fewer non-noisy units left to the model to infer the parameters from.

\subsection{Model selection}
\label{sec:mod_sel}

Following the setup described in Section \ref{sec:sim_set}, by varying $N \in \{300,1500,3000\}$ and adding increasing noise ratios $\tau \in \{0,0.1,0.2\}$, 9 different scenarios have derived for testing the model selection capabilities. We recall that for each scenario and each $N$, 100 data sets have been drawn. Model selection has been performed through BIC, as described in Section \ref{sec:choiceK}. The results are shown in Table \ref{table:BIC_ord}.

For $N = 300$, all the simulated data sets yield a lower BIC for $K$ equal to 2 than 3. However, for larger sample sizes, the model with $K = 3$ is selected for each synthetic data sets in each scenario. The model seems therefore sensitive to sample sizes as small as 300, and seems prone to select a value for $K$ smaller than the actual one for small samples. In this context, it is worth recalling that the BIC is asymptotically consistent. Therefore, one may not be surprised to the fact that for small sample sizes it encounters some issues in selecting the true model.

\begin{table*}[ht]
\begin{adjustwidth}{-1cm}{-1cm}
\centering
\begin{tabular}{|rrrrrrr|rrrrrrr|rrrrrrr|}
  \hline 
\multicolumn{7}{|c|}{Scenario $\tau = 0$} & \multicolumn{7}{|c|}{Scenario $\tau = 0.1$} & \multicolumn{7}{|c|}{Scenario $\tau = 0.2$} \\
\hline
N/K & 1 & 2 &  3  & 4 & 5 & 6 & N/K & 1 & 2 &  3  & 4 & 5 & 6 & N/K & 1 & 2 &  3  & 4 & 5 & 6\\ 
\hline
300 & 0 & 100 &  0  & 0 & 0 & 0 & 300 & 0 & 100 &  0  & 0 & 0 & 0 & 300 & 0 & 100 &  0  & 0 & 0 & 0\\ 
1500 & 0 & 0 & 100 & 0 & 0 & 0 & 1500 & 0 & 0 & 100 & 0 & 0 & 0 & 1500 & 0 & 0 & 100 & 0 & 0 & 0\\
3000 &  0 & 0 &  100 & 0 & 0 & 0 & 3000 &  0 & 0 &  100 & 0 & 0 & 0 & 3000 &  0 & 0 &  100 & 0 & 0 & 0\\
\hline
\end{tabular}
\end{adjustwidth}
\caption{\label{table:BIC_ord} Frequency of selection of each model K by MOM through BIC among the 20 simulated data sets, for increasing N. The actual value for K is 3. Kmeans++ initialization.}
\end{table*}

Looking at the performances in selecting the right $K$ in presence of noise, we can say that overall the model seems able to handle well some noise in the data, provided a sufficient number of remaining non-noisy units to draw its inference from is given. It keeps optimal classification results for units which follow the distributional assumption and selects the correct model even for $\tau = 0.2$.\\
At the same time, the presence of noise makes more extreme the problem of selection of $K$ for small sample size described in the previous paragraph, as the model has even fewer non-noisy units to compute the parameters from.

\subsection{Comparison with competitors}
Finally, we compared the results obtained for the MOM model to the ones given by its continuous version, the Mixture of Matrix-Normals (MMN) (\cite{Viroli2011Oct}), mentioned in Section \ref{sec:relworks}, by treating our ordinal data as continuous ones, as frequently done by practitioners. Moreover, we compared our model against a plain mixture of multivariate normal distributions as well, applied on the vectorized version of the data. To do so, we used the R package \texttt{mclust} (\cite{mclust}).\\
The hyper-parameters of the competitors have been set to be similar to the one of the MOM in terms of convergence and covariance matrix parametrization. Hence, in both cases the stopping rule is given by the absolute difference of two consecutive log-likelihoods being less than $1 \times 10^{-3}$ and the two covariance matrices for MMN and the single one for \texttt{mclust} are fully parametrized.\\
Moreover, we think it is worth mentioning that we tried to perfom the comparison also with the package \texttt{clustMD}, by again running the algorithm on the vectorized version of the data. However, the algorithm was not able to produce any meaningful result. We believe this may be due mainly to the different way the package chooses its thresholds, resulting in computational issues by \texttt{clustMD} for data generated as described in Section \ref{sec:sim_set}.\\

\begin{figure}[ht!]
\centering
\includegraphics[width = 1\linewidth]{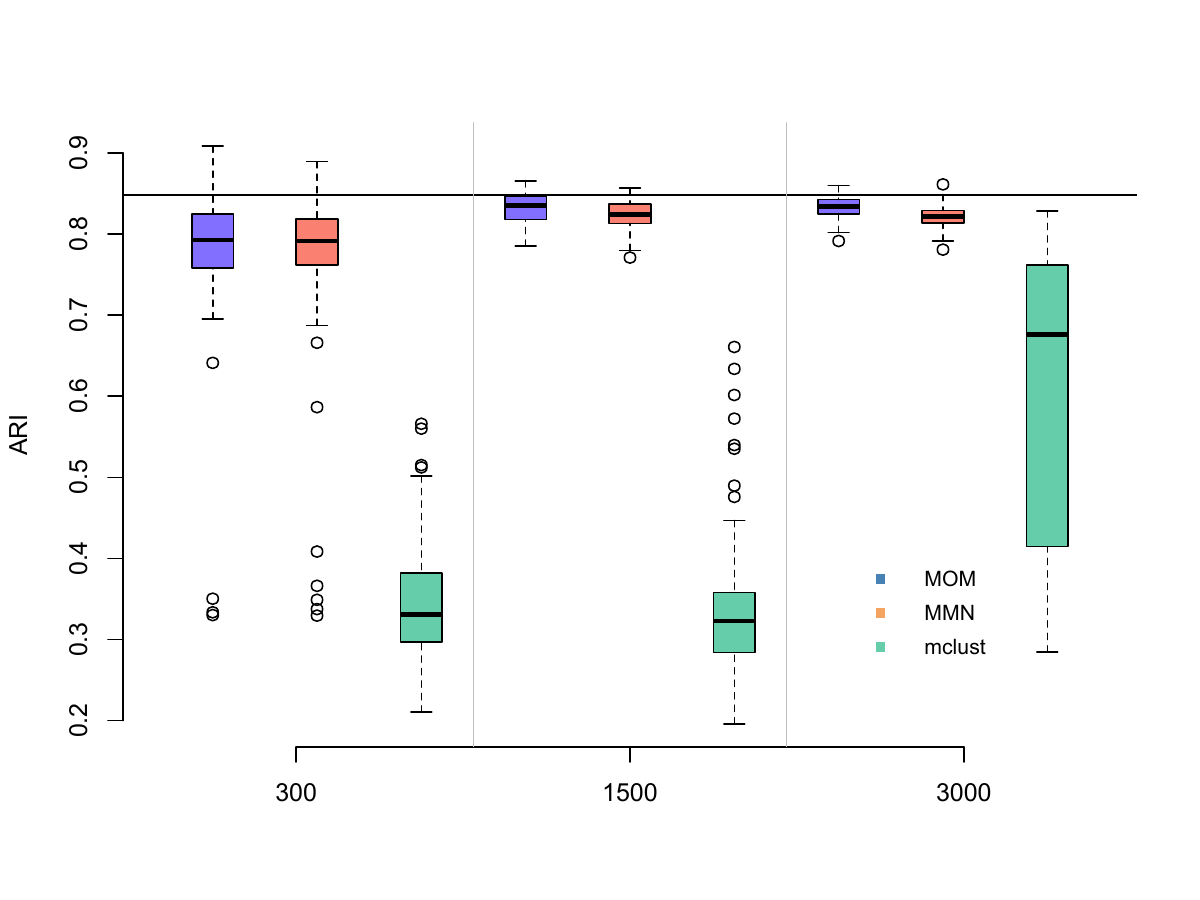}
\caption{ARI for MOM, MMN and mclust. Kmeans++ initialization for MOM and MMN.}
\label{fig:ARI_cont}
\end{figure}

In Figure \ref{fig:ARI_cont} the results for the partitioning task are shown. The difference in the ARI measurement is negligible for $N = 300$ for the two matri-variate model, but increases as $N$ increases. On the other hand, mclust is outperformed consistently.

\begin{figure*}[!ht]
\hspace*{-.2in}
\begin{subfigure}{.5\textwidth}
  \centering
  \includegraphics[width=1\linewidth]{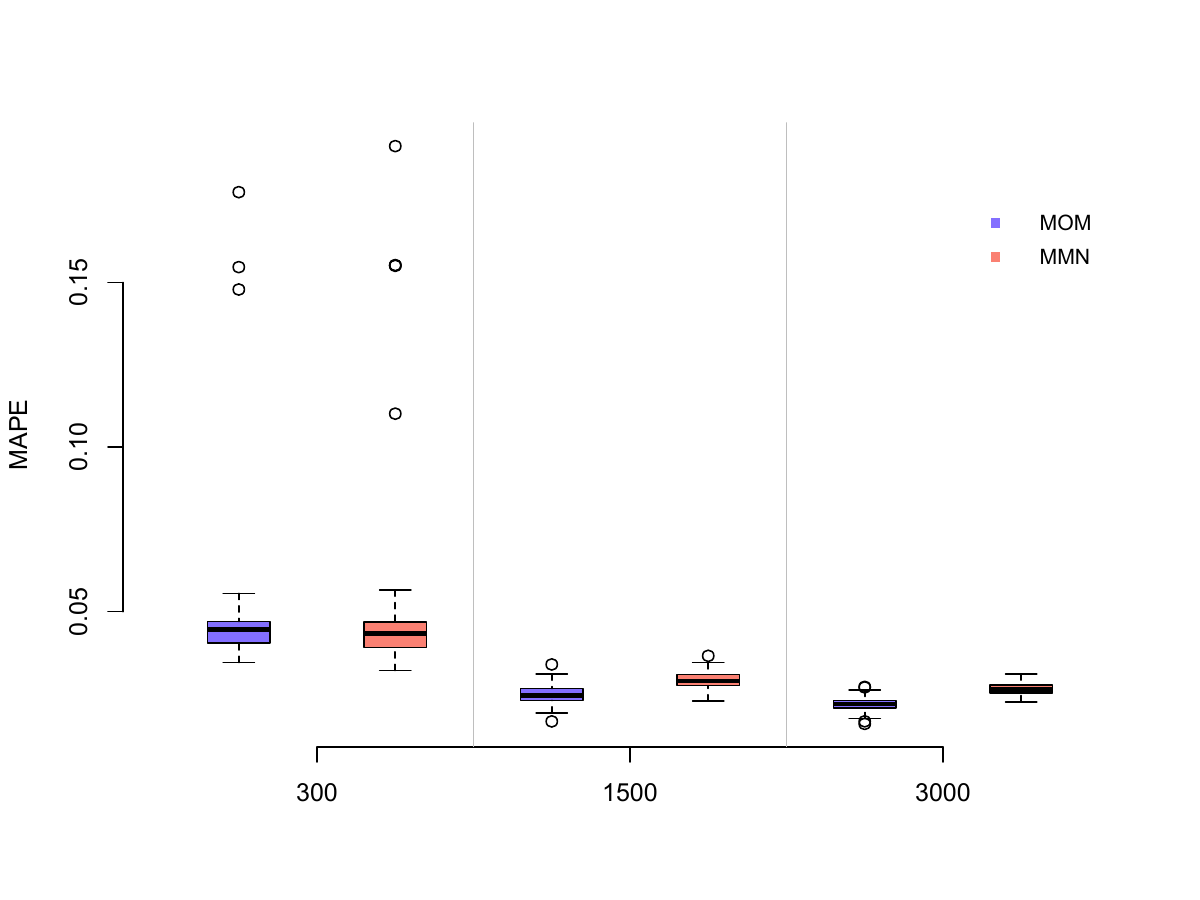}
  \label{fig:MAPE_M}
  \caption{MAPE for $M$}
\end{subfigure}%
\hspace*{.2in}
\begin{subfigure}{.5\textwidth}
  \centering
  \includegraphics[width=1\linewidth]{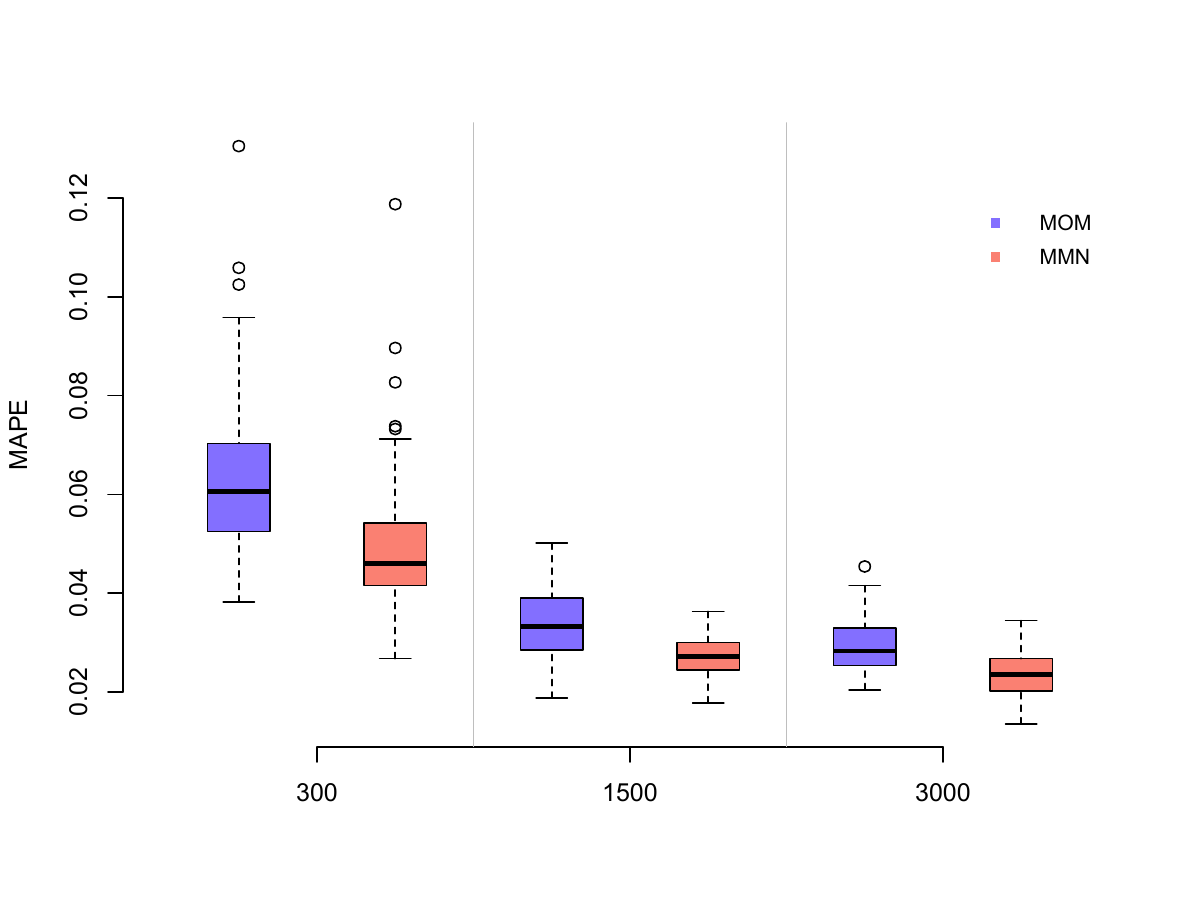}
  \caption{MAPE for $\Phi$}
  \label{fig:MAPE_Phi}
\end{subfigure}
\hspace*{1.5in}
\label{fig:corMMN}
\begin{subfigure}{.5\textwidth}
  \centering
  \includegraphics[width=1\linewidth]{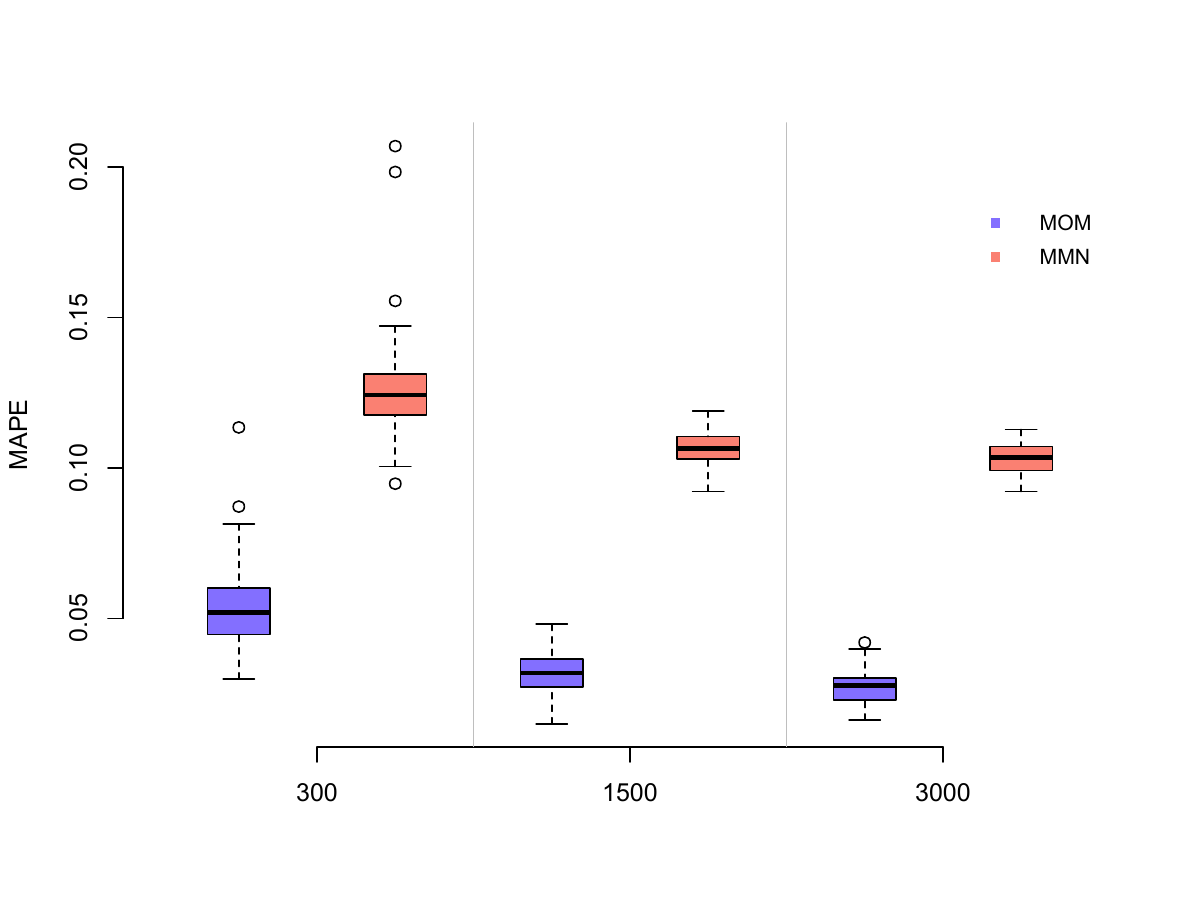}
  \caption{MAPE for $\Sigma$}
  \label{fig:MAPE_Sigma}
\end{subfigure}
\hspace*{.5in}
\caption{MAPE results for parameter matrices. MOM vs MMN. Kmeans++ init.\\ 
Note the difference in the scales.}
\label{fig:MAPE_cont}
\end{figure*}

The difference between MOM and MMN is clearer when comparing the MAPE values for the parameters estimation. As shown in Figure \ref{fig:MAPE_cont}, the distance in error increases as $N$ increases for $M$ and $\Sigma$, but the same does not happens for the diagonal of $\Phi$, for which the MMN method seems to perform better, even if the difference dims as the sample size increases.\\

\section{Real Data}
\label{sec:real}
\subsection{Data}
\label{sec:data}
 After the evaluation of the model through simulations, a real data application concerning preferences for grocery shopping during the Covid-19 pandemic in France \citep{allaz} has been performed. The surveys consists of 78 questions for the first survey (T1), 73 questions for the second (T2) and 55 questions for the remaining three surveys (T3, T4, T5). The answers are mainly on an ordinal scale, and has been conduced at 5 period during the two years of pandemic's intermittent lockdowns to a French sample. The five period at which the surveys has been conduced are: March 26 - April 5, 2020 (beginning of the $1^{\text{st}}$ lockdown); April 30 - May 11, 2020 (end of the $1^{\text{st}}$ lockdown); June 9 - June 16, 2020 (post-lockdown); October 28 - November 9, 2020 (beginning of the $2^{\text{nd}}$ lockdown); March 5 - March 25, 2021 (just before the $3^{\text{rd}}$ lockdown). 
 As part of a preliminary analysis on the data, we have selected 11 questions coming from 3 macro-area of questioning (quoted as Q5, Q8 and Q12). The total number of participants answering  for these 11 questions at each of the 5 surveys is 337. Translated to English, the questions are the following:
\begin{itemize}
    \item Q5: In the last month, you would say that you have preferred in your purchases\ldots
    \begin{itemize}
        \item (1) Seasonal products 
        \item (2) Products "Bio" 
        \item (3) Local products
        \item (4) Fair trade products 
        \item (5) Bulk products (excluding fruit and vegetables) 
    \end{itemize}
    \item Q8: Choose the appropriate answer for each item
    \begin{itemize}
        \item (1) About the foods, you have the impression of wasting
        \item (2) You have paid attention to the expiration dates 
        \item (3) You have prepared anti-waste cooking recipes
    \end{itemize}
    \item Q12: Would you say
    \begin{itemize}
            \item (1) This period is ideal to rethink our way of consuming
            \item (2) This period is ideal to test more environmentally responsible ways of living 
            \item (3) This period is ideal to learn how to consume less 
        \end{itemize}
    \end{itemize}
For each question, the participant have to answer on an ordinal scale 7 levels: for the macro-group Q5 and Q8 the range is from 1 for ``much less than before confinement" to 7 for ``much more than before confinement", while for the macro-group Q12 from 1 for ``high disagreement" to 7 for ``high agreement". In all of the cases the 4th level express some form of ``neutrality".\\
It is worth noticing that the item Q8(1) is an inverse item. As we will see, this will not impact our clustering, as our model is able to handle such items without the need to reverse them, but it is necessary to keep in mind their nature at the moment of interpretation, as it would impact the direction of the correlation with the other items.

So, to sum up, we have N = 337 units for J = 11 variables (questions) and T = 5 times. 

\subsection{Results}
\begin{figure*}[!ht]
%\hspace*{-1.5in}
\begin{subfigure}{.5\textwidth}
  \centering
  \includegraphics[width=1\linewidth]{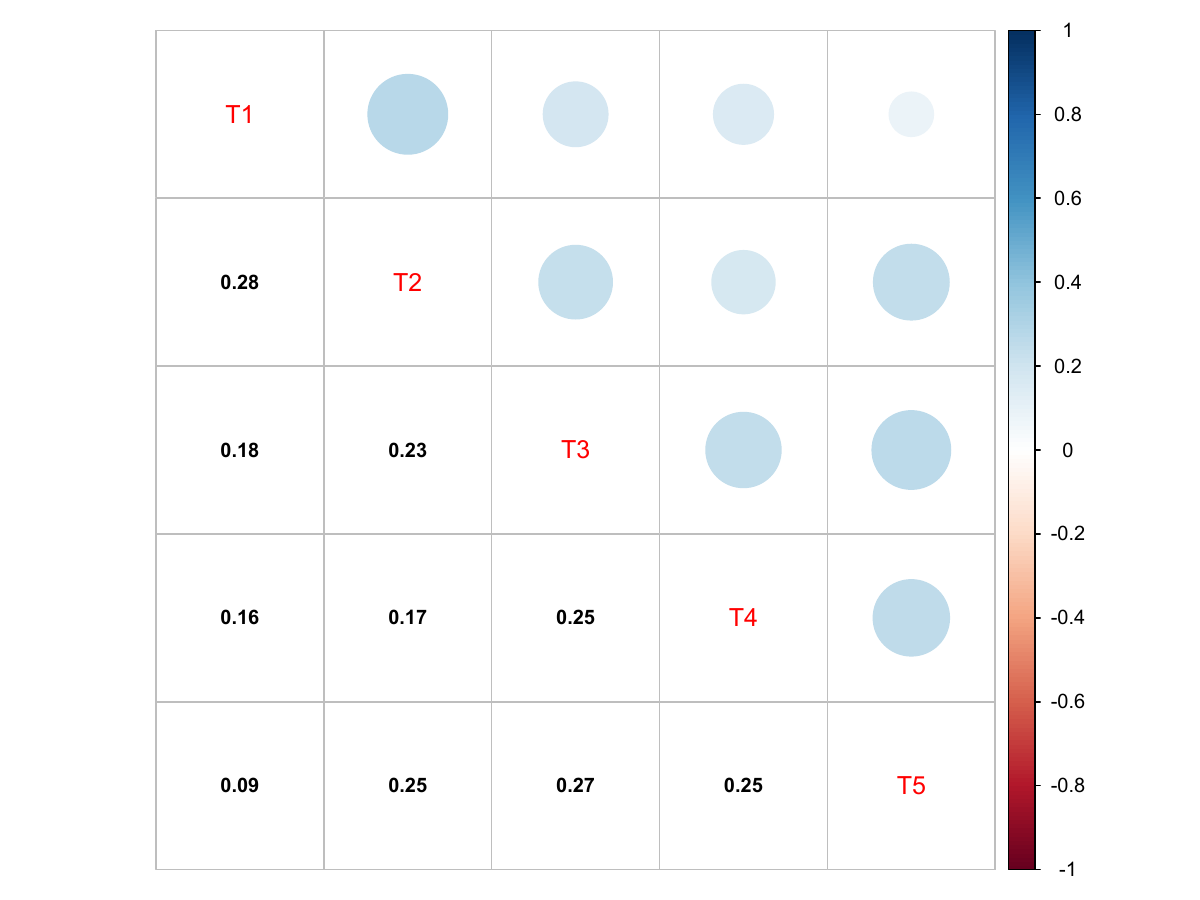}
  \caption{Cluster 1}
  \label{fig:Phi_C1}
\end{subfigure}%
%\hspace*{-.6in}
\begin{subfigure}{.5\textwidth}
  \centering
  \includegraphics[width=1\linewidth]{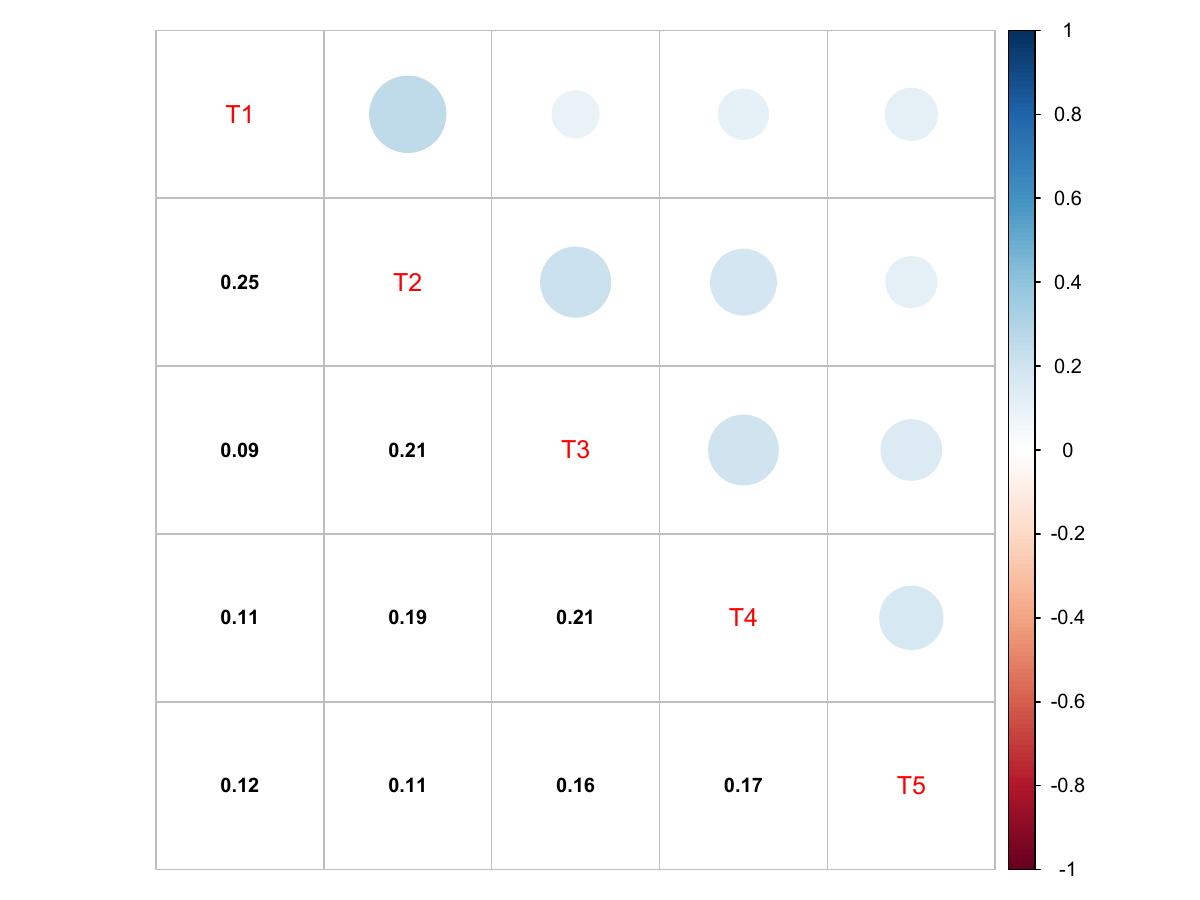}
  \caption{Cluster 2}
  \label{fig:Phi_C2}
\end{subfigure}
\hspace*{1.4in}
\begin{subfigure}{.5\textwidth}
  \centering
  \includegraphics[width=1\linewidth]{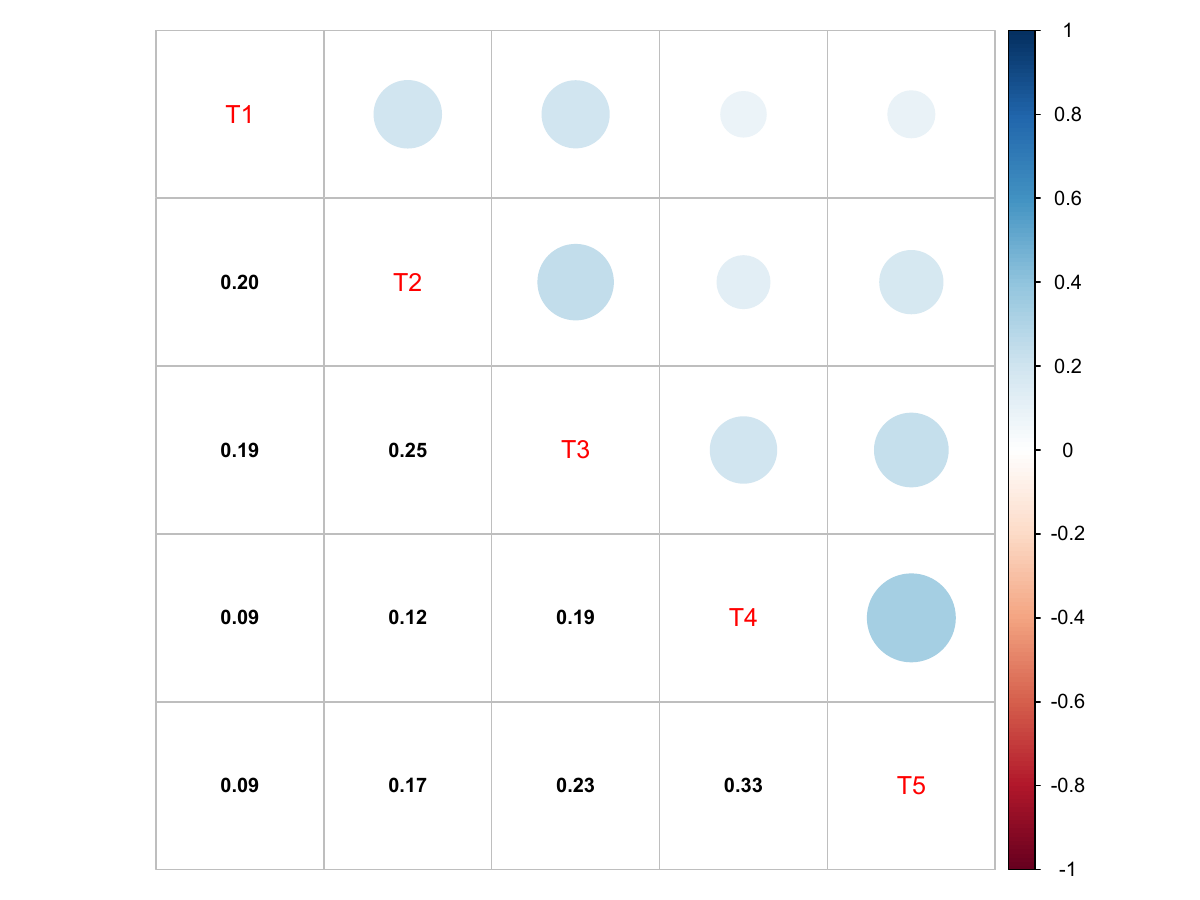}
  \caption{Cluster 3}
  \label{fig:Phi_C3}
\end{subfigure}
\caption{Clusters’ corr-plots among time.}
\label{fig:corrPhi}
\end{figure*}

\begin{figure*}[ht!]
%\hspace*{-1.85in}
\begin{subfigure}{.5\textwidth}
  \centering
  \includegraphics[width=1.1\linewidth]{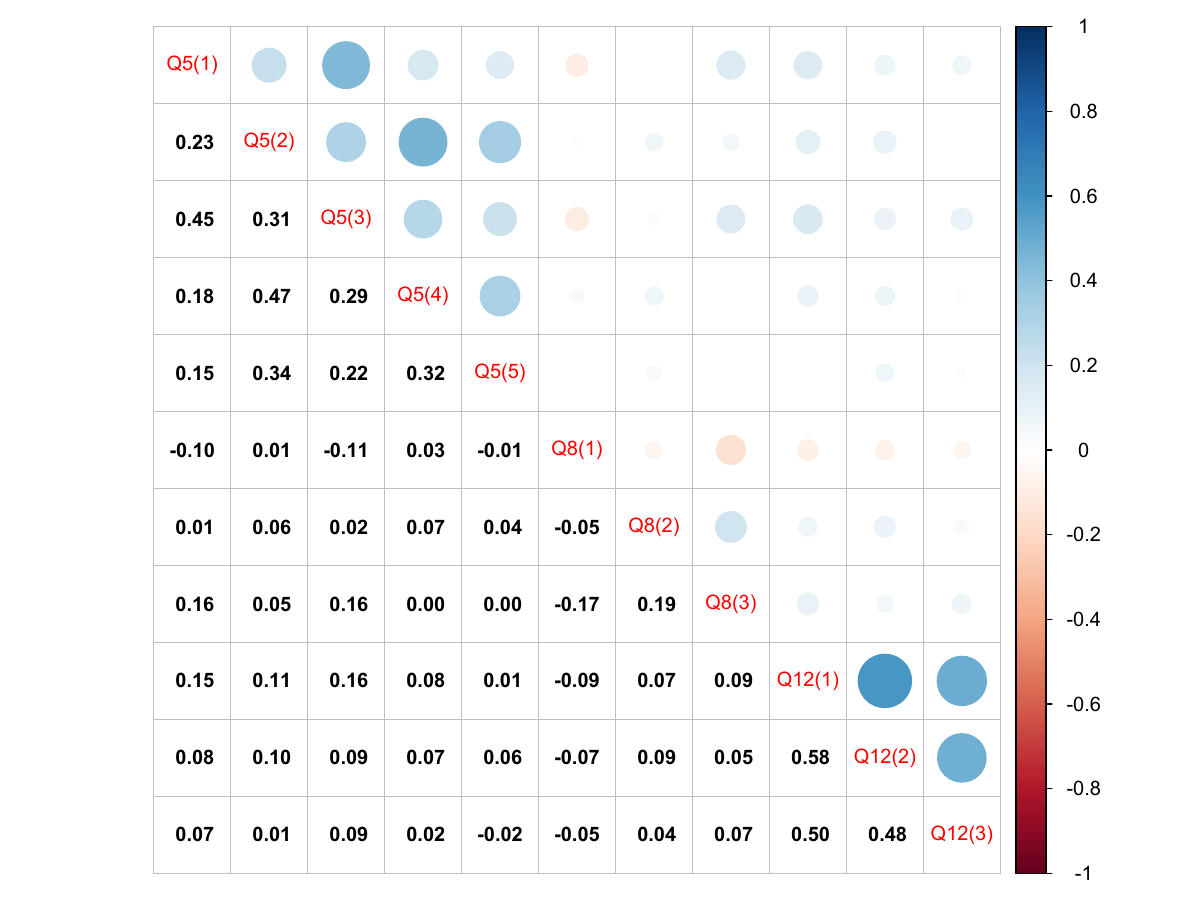}
  \caption{Cluster 1}
  \label{fig:Sigma_C1}
\end{subfigure}%
%\hspace*{-.7in}
\begin{subfigure}{.5\textwidth}
  \centering
  \includegraphics[width=1.1\linewidth]{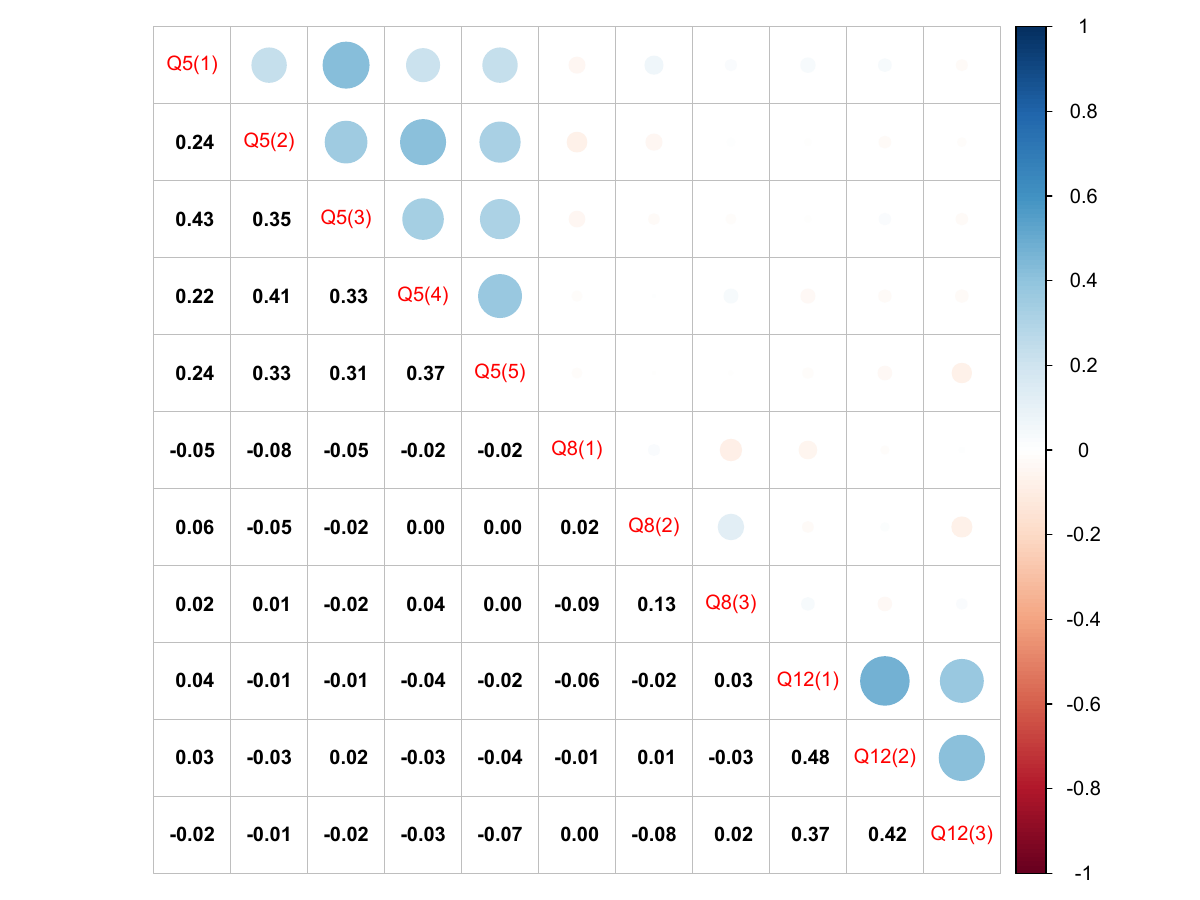}
  \caption{Cluster 2}
  \label{fig:Sigma_C2}
\end{subfigure}
\hspace*{1.4in}
\begin{subfigure}{.5\textwidth}
  \centering
  \includegraphics[width=1.1\linewidth]{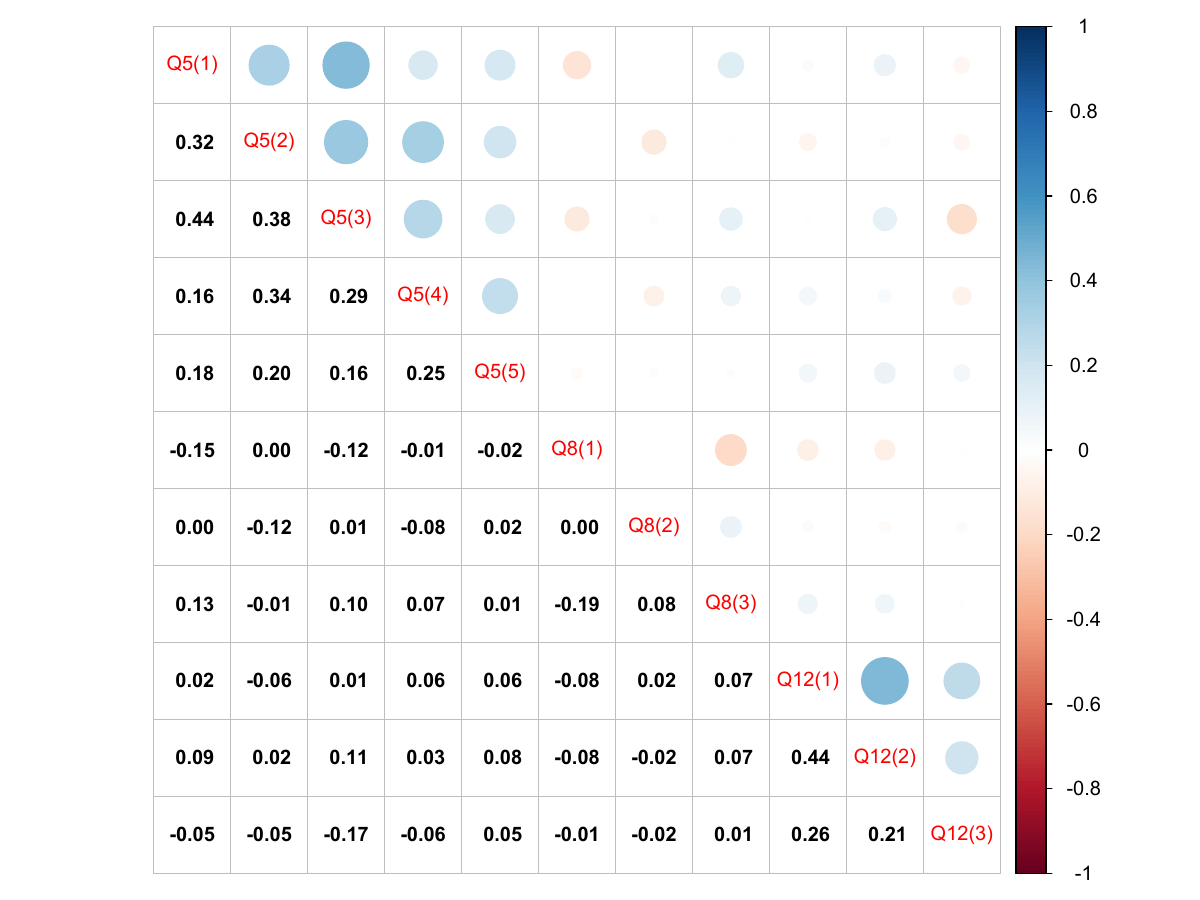}
  \caption{Cluster 3}
  \label{fig:Sigma_C3}
\end{subfigure}
\caption{Clusters’ corr-plots among variables.}
\label{fig:corrSigma}
\end{figure*}

After performing our clustering algorithm with a number of clusters K ranging from 1 to 6 using Kmeans++ initialization, the model with the lowest BIC is with K = 3 (Figure \ref{fig:BIC_realdata_viz}). %Table \ref{table:BIC_realdata}). 
The number of units in first cluster is 124, in the second one they are 149 and in the third 64.
The estimated parameters are reported in Table \ref{table:means} for the mean $M$, Table \ref{table:Phi_cov} for the time covariances $\Phi$ and in Table \ref{table:Sigma_cov} for the variable covariances $\Sigma$. To gain interpretability, covariances matrices have been transformed in correlation matrices in Table \ref{table:Phi} for $\Phi$ and in Table \ref{table:Sigma} for $\Sigma$. In the tables the questions are named using their codes. Moreover, the correlation matrices $\Phi$ and $\Sigma$ are represented by correlation plots in Figures \ref{fig:corrPhi} and \ref{fig:corrSigma}, respectively. 

\begin{figure*}[ht!]
\centering
\includegraphics[width = 1\linewidth]{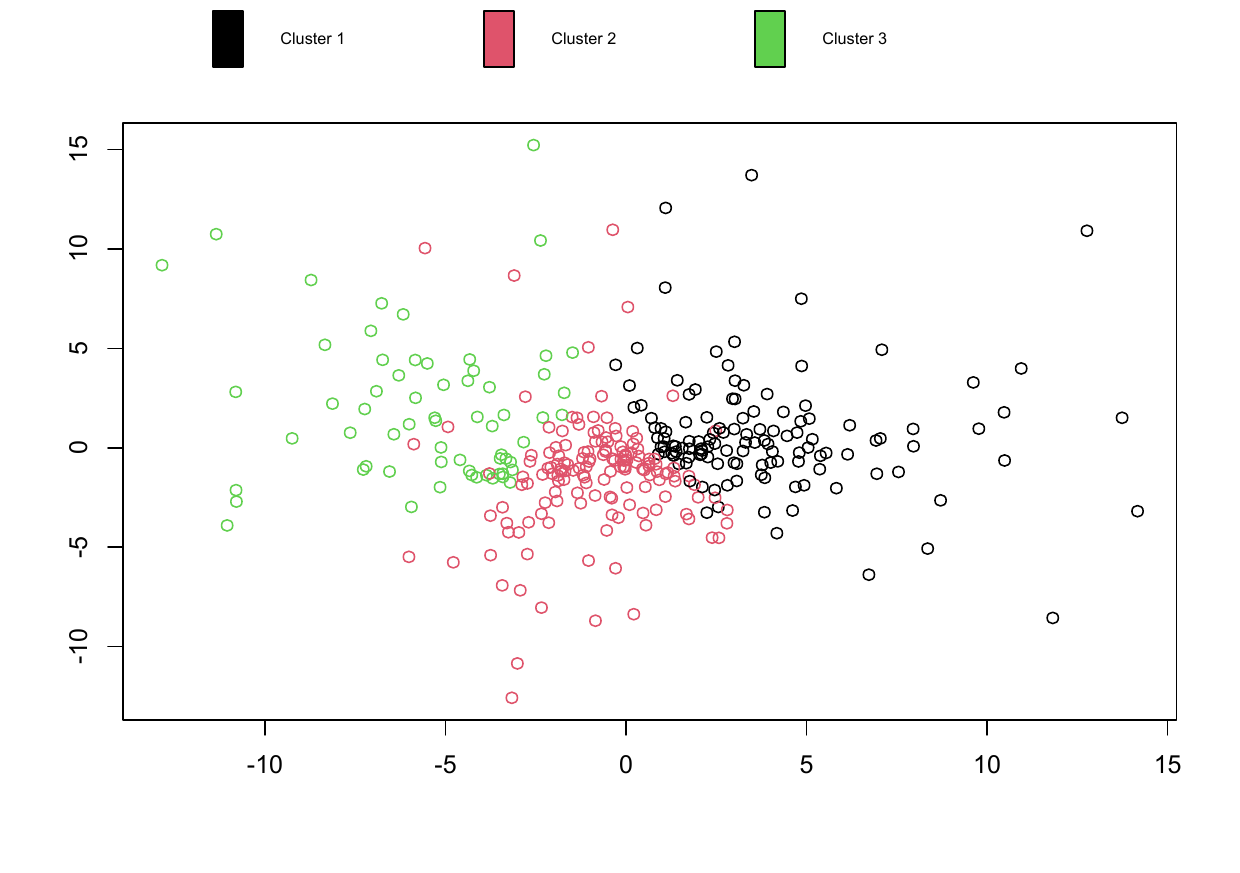}
\caption{Units represented through isoMDS and colored by cluster allocation.}
\label{fig:clust_units}
\end{figure*}

Figure \ref{fig:clust_units} represents the 337 units (individuals) using a non-metric MDS (\cite{Venables2002}), specifically through the function $\mathsf{isoMDS}$ of the R package $\mathsf{MASS}$. In non-metric MDS only the order of dissimilarities is important rather than the amount of dissimilarities, that makes it suitable to be used for ordinal data, as in our case. For this representation, the temporal structure has been discarded and we have transformed our units from $11\times 5$-dimensional matrices to $55$-dimensional vectors. Each individual is represented by a circle whose color depends on its cluster.

\begin{figure*}[ht!]
\centering
\includegraphics[width = 1\linewidth]{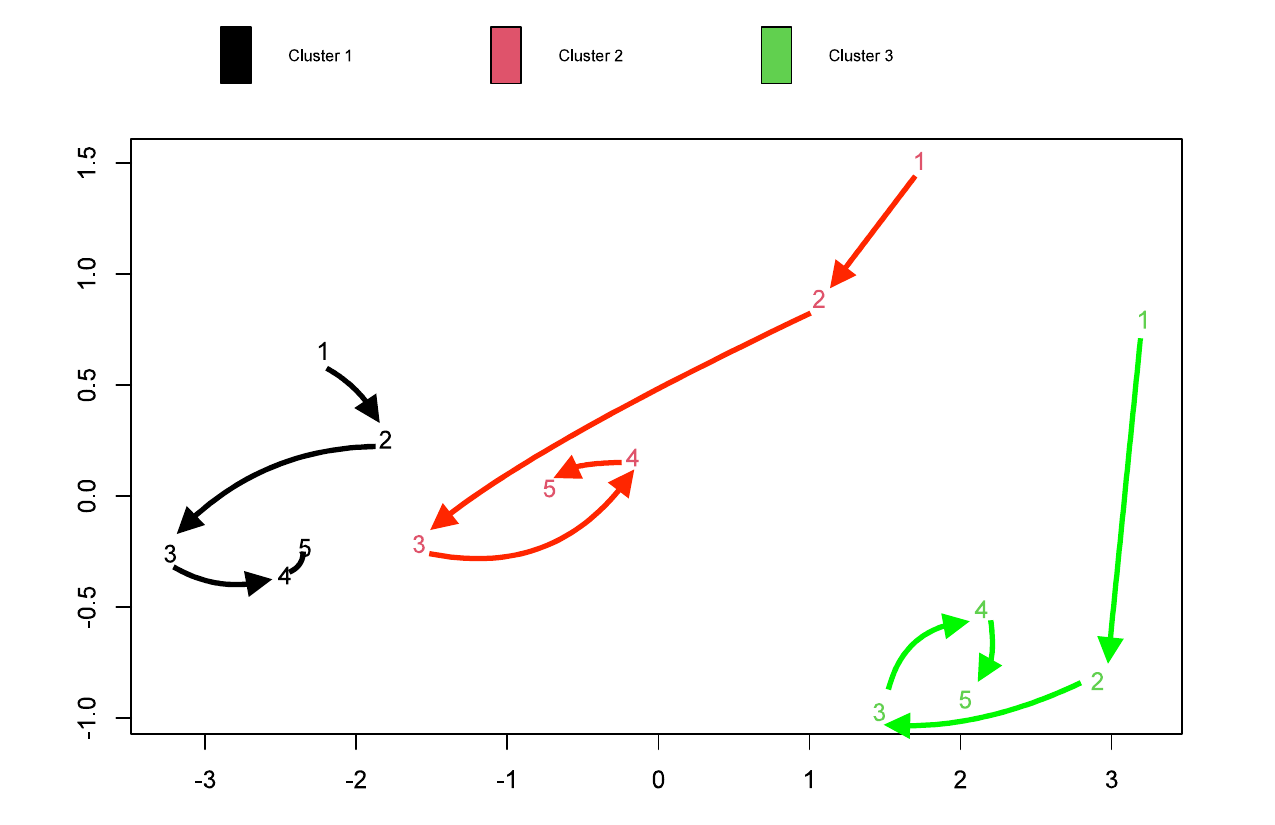}
\caption{Evolution in time of cluster means. Representation through isoMDS.\\
Numbers represent the time and the colors indicate the clusters.} 
\label{fig:clust_ev_cont}
\end{figure*}

Figure \ref{fig:clust_ev_cont} plots, using the same non-metric MDS, the cluster means at each of the 5 times. Such plot allows to visualize the time evolution of each cluster. %A version of the graphic made on discretized mean matrices is shown in Figure \ref{fig:clust_ev_ord}.

\subsection{Interpretation}
 Even if data are represented by means of a dimensionality reduction technique, discarding the temporal structure of data, we can see on Figure \ref{fig:clust_units} that the clusters are well separated. In particular, Cluster 2 is between Cluster 1 and Cluster 3. This fact can be confirmed by looking more finely at Table \ref{table:means}. Moreover, from Figure \ref{fig:clust_ev_cont} it is possible to visualize the comprehensive evolution in time for the clusters means. Indeed, one can see that Cluster 2 and Cluster 3 starts relatively close to one another, but Cluster 2 then evolves and approaches Cluster 1 in T3, to then stabilizing on a more intermediate space. Cluster 1 appears to be the most stable one, moving itself on a confined area of the graph. Cluster 3, despite starting on values close to Cluster 2, evolves differently from the others.

 In the following, we  give a summary description for each cluster and we will try to draw some interpretations. We will start by interpreting Clusters 1 and Cluster 3, which are the most characteristic, to finish with Cluster 2, which could be seen as an intermediary cluster between the other two.
 
\begin{itemize}
\item \textbf{Cluster 1}: 124 units.
    \begin{itemize}
            \item \textbf{Correlation in time}: the cluster is characterized by a fading correlation of T1 with other times and by  generally higher correlations than other clusters, with the exception of just a small rift between T2 and T4.
            \item \textbf{Means}: this is the cluster with overall lowest and most stable mean values, around neutrality level. The only values lower than neutrality are for Q8(1), an inverse item.
            \item \textbf{Correlation among questions}: generally positive correlations or feeble ones, the cluster is mainly characterized by some positive correlations between macro-area Q5 and Q12, and some negative correlations between those areas and Q8(1).
    \end{itemize}
We can characterize Cluster 1 as the cluster with overall neutrality-level and stable means. Indeed, considering that levels range from 1 to 7 as detailed in Section \ref{sec:data}, the values  tend to be around the “neutrality” level, the level coded as 4. Therefore, the cluster is actually a cluster composed by people who were generally neutral with respect to the questions, and did not evolve on this neutrality much during the study period.\\
Looking at Table \ref{table:means}, it is evident that the questions that discriminate the most among the clusters in terms of average level of response are the ones in Q12, the ones regarding rethinking our lifestyle, as they show different average levels for each cluster.
For cluster 1, the average response shows neutrality even in that regard.\\
This cluster is also the ones that has the highest correlations between Q8(3), anti-wasting recipes, and questions in Q5 group. Overall, observing the behaviour of the correlations, seems clear that Cluster 1, despite being the most neutral cluster in terms of average responses, could be defined as the most consistent cluster, since responses that regarding preferences for sustainable grocery shopping are positively correlated with preparations anti-waste recipes and rethinking our way of life.\\
The generally positive correlations among some of the other questions may indicate a certain coherence around the neutrality, given that  preference for a more sustainable grocery shopping is positively correlated to the anti-wasting behaviours and the belief that the pandemic period should inspire a change in the life habits. This signals that the subjects' responses to those topics move likewise within the cluster.\\
In other words, Cluster 1 did not really change its habits (as level 4 means "as before") and appears not to have felt very impacted by the health crisis, as the neutral level on rethinking its way of life may indicate.

\item \textbf{Cluster 3}: 64 units.
    \begin{itemize}
            \item \textbf{Correlation in time}: Cluster 3 seems defined by two correlations blocks; one composed by T1 and T2 and the second by T3,T4 and T5.
            \item \textbf{Means}: with respect to the other clusters, this cluster is characterized by the highest levels for the macro-area Q12 and the lowest values for Q8(1), coherent with the inverse item.
            \item \textbf{Correlation among questions}: the cluster is the most varied one compared to the other clusters. Intra-macro-area correlations are weaker as well. Some noteworthy negative correlation between Q8(2) and Q5(2) and between Q12(3) and Q5(3).
     \end{itemize}
Cluster 3 also has generally neutrality-level values for most of questions belonging to Q5 and Q8 macro-groups throughout the study period, as Cluster 1, despite having some lower values for Q8(1) and some higher ones for Q5(3). The main difference is however in the Q12 macro group, the one we can define as composed by the “rethinking-way-of-life” questions. Cluster 3 has remarkably high values here, meaning that this group of people really found that the pandemic period was stimulating a reflection on our lifestyle. As it turns out, this opinion fades as we advance towards T3 to then re-approach higher levels. It is interesting to observe that T3 corresponds to the beginning of June 2020, that is after the end of the first lockdown, while T4 is at the end of October and beginning of November 2020, after the summer and at the beginning of the second lockdown, and that T5 is in March 2021, when the country was approaching a third lockdown. So, apparently, the second lockdown brings back a reflection on how to live. It seems that people need crises to reflect on their lifestyle.\\
In this cluster we also observe some negative correlations between question Q12(3), concerning less consumption, and Q5(3), which measures the preference for local products, and also between Q8(2), paying attention to expiring dates, and Q5(2) and Q5(4), the preference for ``bio" products and fair trade ones. This may signal that the people composing this cluster who pay more attention to buy ``local" (such as going to the local markets), ``bio" and sustainable fair product may also be the ones who tend to be less concerned regarding consuming less, probably because they already satisfy their concerns by orienting their grocery shopping to more sustainable products. They satisfy their concerns for consuming less by consuming better. 

\item \textbf{Cluster 2}: 149 units.
      \begin{itemize}
            \item \textbf{Correlation in time}: cluster 2 presents notably overall fading correlations in time. 
            \item \textbf{Means}: responses for macro-areas Q5 and Q8 show levels around neutrality, while for macro-area Q12 the levels are middle-high, intermediary between the other two clusters.
            \item \textbf{Correlation among questions}: cluster characterized by generally low correlations among questions of different macro-areas. Some weak negative correlations among Q12(3) and Q5(5) and Q8(2).
\end{itemize}
Cluster 2, as already said, seems composed by subjects whose answers to the questionnaires can be seen as intermediate between the Cluster 1 and Cluster 3. Levels for questions in the Q5 group tend to be lower at the beginning of the inquiry to then have a slight increase over the study period. Questions of the macro-group Q12, that we saw characterized cluster 3 for their high levels in the answers, have an high level for this cluster at the beginning as well, even if not as high as cluster 3. Yet, their value tend towards the ``neutrality" approaching T3, to then have a slight increase. We can think of these subjects as people that highly agreed with changing their way of life at the beginning of the inquiry, to then become more and more disaffected as the strict lockdown period gives way to reestablish a more 'ordinary' way of life. \\
One characteristic of Cluster 2 is that there are not clearly strong correlations outside macro-area blocks, ans even for block Q8 they are not as strong as other clusters. This may indicate heterogeneity in the answers' patterns to the questioners outside the blocks, giving rise not so strong correlations.\\
Some weak negative correlations between Q12(3) and Q5(2) and between Q8(2) and Q5(2) may signal a similar behaviours as in Cluster 3 regarding satisfying their concerns for consuming less by consuming better, even if less pronounced. 
\end{itemize}

Finally, there are some comments to be made about Q8(1) and intra-group correlations.
As said in Section \ref{sec:data}, Q8(1) is an inverse item, and it has indeed negative correlations with other questions. The question asks whether the respondent has the impression to waste. Its negative correlation with questions in Q5 and Q12 group, even if only slightly sometimes, means that people that in general have the impression of wasting food are the ones that report lower values regarding preferences for ``sustainable" grocery shopping and rethinking our way of consuming, while, vice-versa, subjects whose responses have higher values regarding buying local and seasonal product, like people who go to local markets, tend to have a lower impression of wasting, probably because they actively try not to. This indeed connects to the general negative correlation that question Q8(1) and Q8(3) have: as Q8(3) asks whether the respondent has prepared anti-wasting recipes, the negative correlation seems natural.\\
On a final note, it is worth pointing out that the cluster that has the lowest correlations for Q8(1) is Cluster 2, as maybe it contains people that try to buy locally and seasonal but do not arrive at making the effort to prepare anti-waste recipes.
%\clearpage
 
%\clearpage

\section{Conclusions}
In this work we have presented a novel approach for modeling longitudinal ordinal data with unobserved heterogeneity. The model presented does not require the conditional independence assumption and respects the the true nature of ordinal data. The matrix-variate structure allows for a more parsimonious modelling. Also, it can explicitly model the temporal structure and the association among the responses, that can vary among clusters. An EM algorithm to perform inference has been proposed and described. The efficacy of the algorithm has been tested on synthetic data under different sample sizes and different noise ratios.  We proved the goodness of this framework to cluster longitudinal ordinal data and to get cluster that are easy to interpret and to work with even by non-statisticians.

However, the proposed model has some limitations. In this paper we focused only on the simplest structure of matrix-normal distribution. While considerably more parsimonious than a mixture of multivariate normal distributions, the model seems sensitive to small sample sizes, as seen in Section \ref{sec:mod_sel}, since, as the number of clusters increases, the number of parameters to estimate can still became troublesome. To improve this aspect,the covariance matrices can be further decomposed to obtain more flexible and parsimonious models, as done for example in \cite{Anderlucci2015Jun} and in \cite{Sarkar2020Feb}.Besides, by applying a modified Cholesky decomposition on the time-related covariance matrix, one would obtain new matrices whose elements can be interpreted as generalized auto-regressive parameters and innovation variances, as shown by  \cite{Brendan2010}. Moreover, EM algorithm can be leveraged to extend the model to deal with incomplete data under the missing at random (MAR).

Furthermore, typically the data collected in questionnaires are not just ordinal, but rather mixed. Consequently, our final aim is to extend the proposed model to handle longitudinal mixed data, following the frame proposed by \cite{clustmd}.\\ 
Finally, one could as well think of implying, with proper adjustments, different underlying continuous distributions, such as heavy-tailed \citep{Tomarchio2020Dec}, skewed \citep{gallaugher2018finite,Melnykov2018Sep} or t-student \citep{dougru2016finite} distributions to endow the clustering model with different desired properties.

\section*{Acknowledgment}
This work has been realised thanks to the financial support provided by Project IADoc@UdL of the University of Lyon and Université Lumière - Lyon 2 as part of the call for ``doctoral contracts in artificial intelligence 2020" (ANR-20-THIA-0007-01).
We want to thank Agnès François-Lecompte, Morgane Innocent and Dominique Kréziak, co-authors for their work in \cite{allaz} for sharing their data. We would also like to thank Brendan Murphy for his invaluable inputs and support throughout the research process. His insights and expertise were instrumental in shaping the direction of this project.

\clearpage

\begin{appendices}
\setcounter{table}{0}
\renewcommand{\thetable}{A\arabic{table}}
\section{Tables}\label{appendix:Table}
% adjust the L and R margins by 1 inch
\begin{table*}[ht!]
\caption{Clusters' means over time. The estimated parameter $\hat{\pi}$ =  (0.37,0.44,0.19)} \label{table:means}
\begin{adjustwidth}{-1.3cm}{-1cm}
\begin{tabular}{|c|rrrrr|rrrrr|rrrrr|}
\hline 
\multicolumn{1}{|c}{ } & \multicolumn{5}{c|}{Cluster 1} & \multicolumn{5}{|c|}{Cluster 2} & \multicolumn{5}{|c|}{Cluster 3} \\
   \hline
Questions & T1 & T2 & T3 & T4 & T5 & T1 & T2 & T3 & T4 & T5 & T1 & T2 & T3 & T4 & T5 \\
  \hline
Q5(1) & 3.99 & 4.16 & 4.20 & 4.22 & 4.17 &  3.80 & 4.08 & 4.22 & 4.21 & 4.18 &
4.27 & 5.04 & 4.92 & 4.59 & 4.85 \\
Q5(2) & 3.60 & 3.77 & 4.02 & 4.02 & 4.15 &
3.72 & 3.79 & 4.10 & 4.07 & 4.13 &
3.83 & 4.36 & 4.48 & 4.35 & 4.47 \\
Q5(3) & 3.89 & 4.22 & 4.19 & 4.42 & 4.35 &
3.73 & 4.03 & 4.35 & 4.30 & 4.25 &
4.49 & 5.43 & 5.16 & 5.23 & 5.28 \\
Q5(4) & 3.51 & 3.78 & 3.95 & 3.98 & 4.03 &
3.49 & 3.78 & 3.99 & 3.97 & 3.98 &
3.53 & 4.08 & 4.26 & 4.34 & 4.44 \\
Q5(5) & 3.32 & 3.64 & 3.86 & 4.14 & 4.03 &
3.37 & 3.61 & 3.96 & 4.00 & 4.11 &
3.69 & 3.78 & 4.21 & 4.30 & 4.39 \\
Q8(1) & 3.36 & 3.42 & 3.61 & 3.66 & 3.64 & 
3.30 & 3.49 & 3.70 & 3.70 & 3.57 &
2.15 & 2.26 & 2.55 & 3.03 & 2.74 \\
Q8(2) & 4.06 & 4.17 & 4.08 & 4.10 & 4.03 &
4.04 & 4.23 & 3.99 & 4.05 & 3.97 &
4.12 & 4.00 & 4.06 & 4.15 & 4.11 \\
Q8(3) & 4.12 & 4.16 & 4.15 & 4.19 & 4.09 &
4.05 & 4.27 & 4.07 & 4.14 & 4.08 &
4.35 & 4.76 & 4.53 & 4.49 & 4.64 \\
Q12(1) & 4.30 & 4.53 & 3.73 & 4.10 & 4.23 &
6.69 & 6.15 & 4.66 & 5.14 & 4.92 &
7.20 & 7.10 & 6.36 & 6.76 & 6.59 \\
Q12(2) & 4.13 & 4.50 & 3.53 & 3.94 & 4.15 &
6.69 & 6.38 & 4.49 & 5.56 & 5.18 &
7.22 & 6.93 & 6.08 & 6.61 & 6.65 \\
Q12(3) & 4.38 & 4.41 & 3.67 & 4.10 & 4.02 &
6.49 & 6.07 & 4.70 & 5.69 & 5.29 &
7.32 & 6.72 & 6.04 & 6.48 & 6.24\\
\hline
\end{tabular}
\end{adjustwidth}
\end{table*}
\break 

\begin{table*}[ht!]
\centering
\caption{Clusters' time correlation}
\label{table:Phi}
\begin{adjustwidth}{-1cm}{-1cm}
\begin{tabular}{|c|rrrrr|rrrrr|rrrrr|}
\hline 
\multicolumn{1}{|c}{} & \multicolumn{5}{c|}{Cluster 1} & \multicolumn{5}{|c|}{Cluster 2} & \multicolumn{5}{|c|}{Cluster 3} \\
 \hline
 T / T & T1 & T2 & T3 & T4 & T5 & T1 & T2 & T3 & T4 & T5 & T1 & T2 & T3 & T4 & T5\\
\hline
T1 & 1.00 & 0.28 & 0.18 & 0.16 & 0.09 &
1.00 & 0.25 & 0.09 & 0.11 & 0.12 &
1.00 & 0.20 & 0.19 & 0.09 & 0.09 \\
T2 & 0.28 & 1.00 & 0.23 & 0.17 & 0.25 &
0.25 & 1.00 & 0.21 & 0.19 & 0.11 &
0.20 & 1.00 & 0.25 & 0.12 & 0.17 \\
T3 & 0.18 & 0.23 & 1.00 & 0.25 & 0.27 &
0.09 & 0.21 & 1.00 & 0.21 & 0.16 &
0.19 & 0.25 & 1.00 & 0.19 & 0.23 \\
T4 & 0.16 & 0.17 & 0.25 & 1.00 & 0.25 &
0.11 & 0.19 & 0.21 & 1.00 & 0.17 &
0.09 & 0.12 & 0.19 & 1.00 & 0.33 \\
T5 & 0.09 & 0.25 & 0.27 & 0.25 & 1.00 &
0.12 & 0.11 & 0.16 & 0.17 & 1.00 &
0.09 & 0.17 & 0.23 & 0.33 & 1.00\\
\hline
\end{tabular}
\end{adjustwidth}
\end{table*}

\begin{table*}[ht!]
\centering
\caption{Clusters' time covariances}
\label{table:Phi_cov}
\begin{adjustwidth}{-1cm}{-1cm}
\begin{tabular}{|c|rrrrr|rrrrr|rrrrr|}
\hline 
\multicolumn{1}{|c}{} & \multicolumn{5}{c|}{Cluster 1} & \multicolumn{5}{|c|}{Cluster 2} & \multicolumn{5}{|c|}{Cluster 3} \\
 \hline
T / T  & T1 & T2 & T3 & T4 & T5 & T1 & T2 & T3 & T4 & T5 & T1 & T2 & T3 & T4 & T5 \\
\hline
T1 & 1.34 & 0.36 & 0.20 & 0.17 & 0.09 &
1.17 & 0.30 & 0.09 & 0.10 & 0.12 &
1.50 & 0.32 & 0.27 & 0.13 & 0.14 \\
T2 & 0.36 & 1.25 & 0.25 & 0.18 & 0.26 &
0.30 & 1.22 & 0.21 & 0.19 & 0.11 &
0.32 & 1.78 & 0.38 & 0.19 & 0.28 \\
T3 & 0.20 & 0.25 & 0.88 & 0.21 & 0.23 &
0.09 & 0.21 & 0.84 & 0.17 & 0.13 &
0.27 & 0.38 & 1.33 & 0.26 & 0.33 \\
T4 & 0.17 & 0.18 & 0.21 & 0.87 & 0.22 &
0.10 & 0.19 & 0.17 & 0.81 & 0.14 &
 0.13 & 0.19 & 0.26 & 1.42 & 0.49 \\
T5 & 0.09 & 0.26 & 0.23 & 0.22 & 0.85 &
0.12 & 0.11 & 0.13 & 0.14 & 0.86 &
0.14 & 0.28 & 0.33 & 0.49 & 1.48\\
\hline
\end{tabular}
\end{adjustwidth}
\end{table*}

\begin{table*}[ht!]
\centering
\caption{Clusters' variables correlation}
\label{table:Sigma}
\begin{adjustwidth}{-1cm}{-1cm}
\begin{tabular}{|crrrrrrrrrrr|}
\hline
\multicolumn{12}{|c|}{\textbf{Cluster 1}}\\
\hline
J / J & Q5(1) & Q5(2) & Q5(3) & Q5(4) & Q5(5) & Q8(1) & Q8(2) & Q8(3) & Q12(1) & Q12(2) & Q12(3) \\
\hline
Q5(1) & 1.00 & 0.23 & 0.45 & 0.18 & 0.15 & -0.10 & 0.01 & 0.16 & 0.15 & 0.08 & 0.07 \\ 
  Q5(2) & 0.23 & 1.00 & 0.31 & 0.47 & 0.34 & 0.01 & 0.06 & 0.05 & 0.11 & 0.10 & 0.01 \\ 
  Q5(3) & 0.45 & 0.31 & 1.00 & 0.29 & 0.22 & -0.11 & 0.02 & 0.16 & 0.16 & 0.09 & 0.09 \\ 
  Q5(4) & 0.18 & 0.47 & 0.29 & 1.00 & 0.32 & 0.03 & 0.07 & -0.00 & 0.08 & 0.07 & 0.02 \\ 
  Q5(5) & 0.15 & 0.34 & 0.22 & 0.32 & 1.00 & -0.01 & 0.04 & -0.00 & 0.01 & 0.06 & -0.02 \\ 
  Q8(1) & -0.10 & 0.01 & -0.11 & 0.03 & -0.01 & 1.00 & -0.05 & -0.17 & -0.09 & -0.07 & -0.05 \\ 
  Q8(2) & 0.01 & 0.06 & 0.02 & 0.07 & 0.04 & -0.05 & 1.00 & 0.19 & 0.07 & 0.09 & 0.04 \\ 
  Q8(3) & 0.16 & 0.05 & 0.16 & -0.00 & -0.00 & -0.17 & 0.19 & 1.00 & 0.09 & 0.05 & 0.07 \\ 
  Q12(1) & 0.15 & 0.11 & 0.16 & 0.08 & 0.01 & -0.09 & 0.07 & 0.09 & 1.00 & 0.58 & 0.50 \\ 
  Q12(2) & 0.08 & 0.10 & 0.09 & 0.07 & 0.06 & -0.07 & 0.09 & 0.05 & 0.58 & 1.00 & 0.48 \\ 
  Q12(3) & 0.07 & 0.01 & 0.09 & 0.02 & -0.02 & -0.05 & 0.04 & 0.07 & 0.50 & 0.48 & 1.00 \\ 
%\space
  \hline
\hline
\multicolumn{12}{|c|}{\textbf{Cluster 2}}\\
 \hline
J / J & Q5(1) & Q5(2) & Q5(3) & Q5(4) & Q5(5) & Q8(1) & Q8(2) & Q8(3) & Q12(1) & Q12(2) & Q12(3) \\ 
 \hline
Q5(1) & 1.00 & 0.24 & 0.43 & 0.22 & 0.24 & -0.05 & 0.06 & 0.02 & 0.04 & 0.03 & -0.02 \\ 
Q5(2) & 0.24 & 1.00 & 0.35 & 0.41 & 0.33 & -0.08 & -0.05 & 0.01 & -0.01 & -0.03 & -0.01 \\ 
Q5(3) & 0.43 & 0.35 & 1.00 & 0.33 & 0.31 & -0.05 & -0.02 & -0.02 & -0.01 & 0.02 & -0.02 \\ 
Q5(4) & 0.22 & 0.41 & 0.33 & 1.00 & 0.37 & -0.02 & 0.00 & 0.04 & -0.04 & -0.03 & -0.03 \\ 
Q5(5) & 0.24 & 0.33 & 0.31 & 0.37 & 1.00 & -0.02 & -0.00 & -0.00 & -0.02 & -0.04 & -0.07 \\ 
Q8(1) & -0.05 & -0.08 & -0.05 & -0.02 & -0.02 & 1.00 & 0.02 & -0.09 & -0.06 & -0.01 & 0.00 \\ 
Q8(2) & 0.06 & -0.05 & -0.02 & 0.00 & -0.00 & 0.02 & 1.00 & 0.13 & -0.02 & 0.01 & -0.08 \\ 
Q8(3) & 0.02 & 0.01 & -0.02 & 0.04 & -0.00 & -0.09 & 0.13 & 1.00 & 0.03 & -0.03 & 0.02 \\ 
Q12(1) & 0.04 & -0.01 & -0.01 & -0.04 & -0.02 & -0.06 & -0.02 & 0.03 & 1.00 & 0.48 & 0.37 \\ 
Q12(2) & 0.03 & -0.03 & 0.02 & -0.03 & -0.04 & -0.01 & 0.01 & -0.03 & 0.48 & 1.00 & 0.42 \\ 
Q12(3) & -0.02 & -0.01 & -0.02 & -0.03 & -0.07 & 0.00 & -0.08 & 0.02 & 0.37 & 0.42 & 1.00 \\ 
\hline
\hline
\multicolumn{12}{|c|}{\textbf{Cluster 3}}\\
\hline
J / J & Q5(1) & Q5(2) & Q5(3) & Q5(4) & Q5(5) & Q8(1) & Q8(2) & Q8(3) & Q12(1) & Q12(2) & Q12(3) \\ 
\hline
Q5(1) & 1.00 & 0.32 & 0.44 & 0.16 & 0.18 & -0.15 & -0.00 & 0.13 & 0.02 & 0.09 & -0.05 \\ 
Q5(2) & 0.32 & 1.00 & 0.38 & 0.34 & 0.20 & -0.00 & -0.12 & -0.01 & -0.06 & 0.02 & -0.05 \\ 
Q5(3) & 0.44 & 0.38 & 1.00 & 0.29 & 0.16 & -0.12 & 0.01 & 0.10 & 0.01 & 0.11 & -0.17 \\ 
Q5(4) & 0.16 & 0.34 & 0.29 & 1.00 & 0.25 & -0.01 & -0.08 & 0.07 & 0.06 & 0.03 & -0.06 \\ 
Q5(5) & 0.18 & 0.20 & 0.16 & 0.25 & 1.00 & -0.02 & 0.02 & 0.01 & 0.06 & 0.08 & 0.05 \\ 
Q8(1) & -0.15 & -0.00 & -0.12 & -0.01 & -0.02 & 1.00 & 0.00 & -0.19 & -0.08 & -0.08 & -0.01 \\ 
Q8(2) & -0.00 & -0.12 & 0.01 & -0.08 & 0.02 & 0.00 & 1.00 & 0.08 & 0.02 & -0.02 & -0.02 \\ 
Q8(3) & 0.13 & -0.01 & 0.10 & 0.07 & 0.01 & -0.19 & 0.08 & 1.00 & 0.07 & 0.07 & 0.01 \\ 
Q12(1) & 0.02 & -0.06 & 0.01 & 0.06 & 0.06 & -0.08 & 0.02 & 0.07 & 1.00 & 0.44 & 0.26 \\ 
Q12(2) & 0.09 & 0.02 & 0.11 & 0.03 & 0.08 & -0.08 & -0.02 & 0.07 & 0.44 & 1.00 & 0.21 \\ 
Q12(3) & -0.05 & -0.05 & -0.17 & -0.06 & 0.05 & -0.01 & -0.02 & 0.01 & 0.26 & 0.21 & 1.00 \\
\hline
\end{tabular}
\end{adjustwidth}
\end{table*}
\clearpage

\begin{table*}
\centering
\caption{Clusters' variables covariances}
\label{table:Sigma_cov}
\begin{adjustwidth}{-1cm}{-1cm}
\begin{tabular}{|crrrrrrrrrrr|}
\hline
\multicolumn{12}{|c|}{\textbf{Cluster 1}}\\
\hline
J / J & Q5(1) & Q5(2) & Q5(3) & Q5(4) & Q5(5) & Q8(1) & Q8(2) & Q8(3) & Q12(1) & Q12(2) & Q12(3) \\
\hline
Q5(1) & 0.58 & 0.14 & 0.31 & 0.10 & 0.10 & -0.06 & 0.00 & 0.08 & 0.15 & 0.08 & 0.07 \\ 
Q5(2) & 0.14 & 0.62 & 0.22 & 0.28 & 0.25 & 0.01 & 0.03 & 0.03 & 0.11 & 0.10 & 0.01 \\ 
Q5(3) & 0.31 & 0.22 & 0.84 & 0.20 & 0.18 & -0.08 & 0.01 & 0.09 & 0.19 & 0.11 & 0.12 \\ 
Q5(4) & 0.10 & 0.28 & 0.20 & 0.56 & 0.22 & 0.02 & 0.03 & -0.00 & 0.08 & 0.07 & 0.02 \\ 
Q5(5) & 0.10 & 0.25 & 0.18 & 0.22 & 0.83 & -0.01 & 0.02 & -0.00 & 0.01 & 0.08 & -0.02 \\ 
Q8(1) & -0.06 & 0.01 & -0.08 & 0.02 & -0.01 & 0.62 & -0.03 & -0.09 & -0.09 & -0.08 & -0.06 \\ 
Q8(2) & 0.00 & 0.03 & 0.01 & 0.03 & 0.02 & -0.03 & 0.45 & 0.09 & 0.06 & 0.08 & 0.03 \\ 
Q8(3) & 0.08 & 0.03 & 0.09 & -0.00 & -0.00 & -0.09 & 0.09 & 0.43 & 0.08 & 0.05 & 0.07 \\ 
Q12(1) & 0.15 & 0.11 & 0.19 & 0.08 & 0.01 & -0.09 & 0.06 & 0.08 & 1.66 & 1.01 & 0.91 \\ 
Q12(2) & 0.08 & 0.10 & 0.11 & 0.07 & 0.08 & -0.08 & 0.08 & 0.05 & 1.01 & 1.79 & 0.91 \\ 
Q12(3) & 0.07 & 0.01 & 0.12 & 0.02 & -0.02 & -0.06 & 0.03 & 0.07 & 0.91 & 0.91 & 2.00 \\ 
\hline
\hline
\multicolumn{12}{|c|}{\textbf{Cluster 2}}\\
\hline
J / J & Q5(1) & Q5(2) & Q5(3) & Q5(4) & Q5(5) & Q8(1) & Q8(2) & Q8(3) & Q12(1) & Q12(2) & Q12(3) \\ 
 \hline
Q5(1) & 0.55 & 0.13 & 0.30 & 0.12 & 0.16 & -0.03 & 0.03 & 0.01 & 0.04 & 0.03 & -0.02 \\ 
Q5(2) & 0.13 & 0.58 & 0.25 & 0.23 & 0.23 & -0.05 & -0.02 & 0.01 & -0.01 & -0.02 & -0.01 \\ 
Q5(3) & 0.30 & 0.25 & 0.89 & 0.23 & 0.27 & -0.04 & -0.01 & -0.01 & -0.01 & 0.03 & -0.03 \\ 
Q5(4) & 0.12 & 0.23 & 0.23 & 0.56 & 0.25 & -0.01 & 0.00 & 0.02 & -0.03 & -0.03 & -0.03 \\ 
Q5(5) & 0.16 & 0.23 & 0.27 & 0.25 & 0.83 & -0.01 & -0.00 & -0.00 & -0.02 & -0.04 & -0.09 \\ 
Q8(1) & -0.03 & -0.05 & -0.04 & -0.01  & -0.01 & 0.79 & 0.01 & -0.05 & -0.07 & -0.01 & 0.00 \\ 
Q8(2) & 0.03 & -0.02 & -0.01 & 0.00 & -0.00 & 0.01 & 0.42 & 0.06 & -0.02 & 0.01 & -0.07 \\ 
Q8(3) & 0.01 & 0.01 & -0.01 & 0.02 & -0.00 & -0.05 & 0.06 & 0.44 & 0.02 & -0.03 & 0.02 \\ 
Q12(1) & 0.04 & -0.01 & -0.01 & -0.03 & -0.02 & -0.07 & -0.02 & 0.02 & 1.52 & 0.73 & 0.63 \\ 
Q12(2) & 0.03 & -0.02 & 0.03 & -0.03 & -0.04 & -0.01 & 0.01 & -0.03 & 0.73 & 1.53 & 0.71 \\ 
Q12(3) & -0.02 & -0.01 & -0.03 & -0.03 & -0.09 & 0.00 & -0.07 & 0.02 & 0.63 & 0.71 & 1.87 \\
\hline
\hline
\multicolumn{12}{|c|}{\textbf{Cluster 3}}\\
\hline
J / J & Q5(1) & Q5(2) & Q5(3) & Q5(4) & Q5(5) & Q8(1) & Q8(2) & Q8(3) & Q12(1) & Q12(2) & Q12(3) \\ 
\hline
Q5(1) & 0.90 & 0.26 & 0.42 & 0.13 & 0.16 & -0.14 & -0.00 & 0.12 & 0.02 & 0.08 & -0.05 \\ 
Q5(2) & 0.26 & 0.74 & 0.33 & 0.24 & 0.16 & -0.00 & -0.09 & -0.00 & -0.05 & 0.02 & -0.05 \\ 
Q5(3) & 0.42 & 0.33 & 1.01 & 0.24 & 0.15 & -0.12 & 0.01 & 0.10 & 0.01 & 0.11 & -0.19 \\ 
Q5(4) & 0.13 & 0.24 & 0.24 & 0.68 & 0.19 & -0.00 & -0.06 & 0.06 & 0.05 & 0.03 & -0.06 \\ 
Q5(5) & 0.16 & 0.16 & 0.15 & 0.19 & 0.84 & -0.02 & 0.01 & 0.01 & 0.06 & 0.08 & 0.05 \\ 
Q8(1) & -0.14 & -0.00 & -0.12 & -0.00 & -0.02 & 1.00 & 0.00 & -0.18 & -0.08 & -0.08 & -0.01 \\ 
Q8(2) & -0.00 & -0.09 & 0.01 & -0.06 & 0.01 & 0.00 & 0.87 & 0.08 & 0.02 & -0.02 & -0.02 \\ 
Q8(3) & 0.12 & -0.00 & 0.10 & 0.06 & 0.01 & -0.18 & 0.08 & 0.91 & 0.07 & 0.06 & 0.01 \\ 
Q12(1) & 0.02 & -0.05 & 0.01 & 0.05 & 0.06 & -0.08 & 0.02 & 0.07 & 1.05 & 0.46 & 0.30 \\ 
Q12(2) & 0.08 & 0.02 & 0.11 & 0.03 & 0.08 & -0.08 & -0.02 & 0.06 & 0.46 & 1.00 & 0.24 \\ 
Q12(3) & -0.05 & -0.05 & -0.19 & -0.06 & 0.05 & -0.01 & -0.02 & 0.01 & 0.30 & 0.24 & 1.26 \\ 
\hline
\end{tabular}
\end{adjustwidth}
\end{table*}

\clearpage

\section{Figures}
\label{appendix:Figures}

\setcounter{figure}{0}
\renewcommand{\thefigure}{B\arabic{figure}}

\begin{figure*}[!ht]
\centering
\includegraphics[scale = .8]{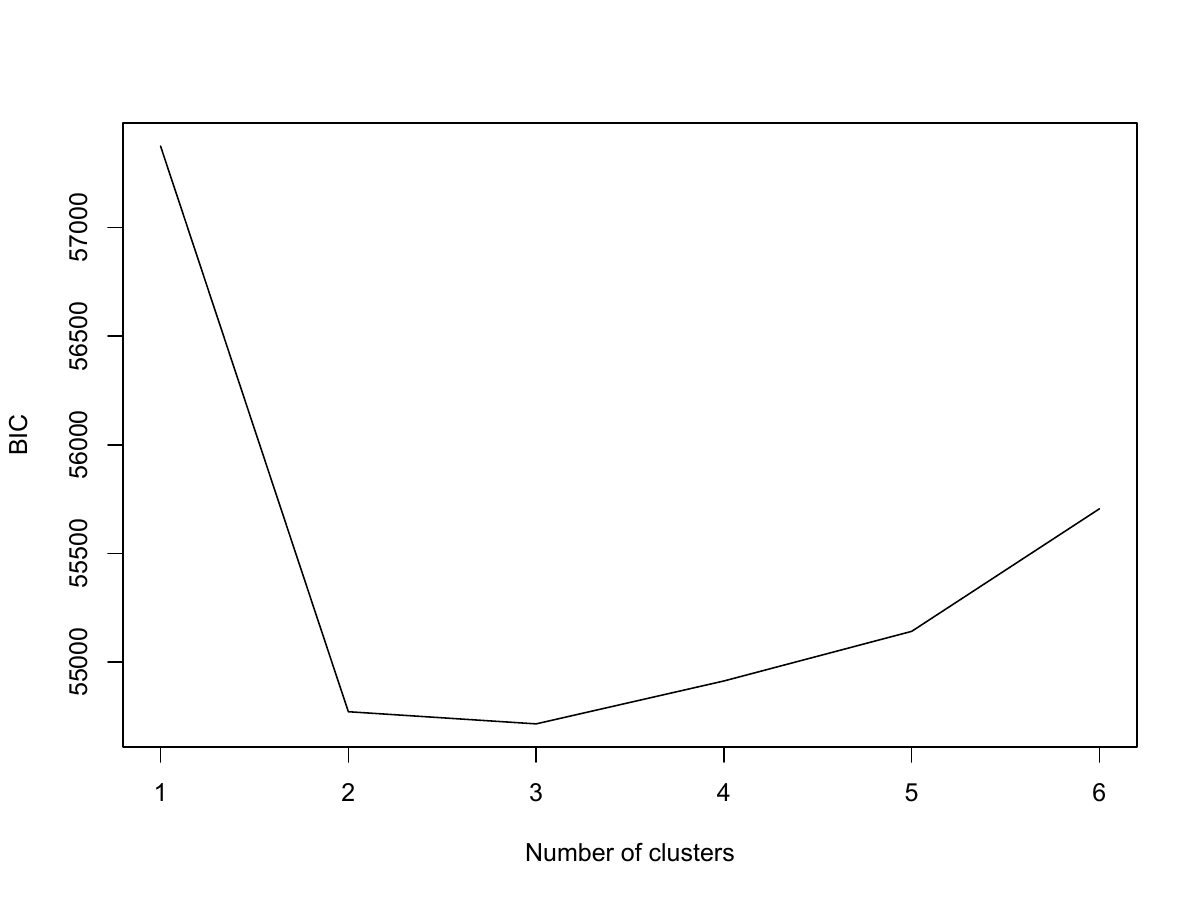}
\caption{Visualization of BIC for K as results of application on real data. Kmeans++ initialization.}
\label{fig:BIC_realdata_viz}
\end{figure*}

\end{appendices}

\clearpage

\printbibliography

\end{document}